\documentclass[a4paper]{article}
\usepackage[latin1]{inputenc}
\usepackage[T1]{fontenc}
\usepackage[english]{babel}
\usepackage{amsmath}
\usepackage{amssymb,amsfonts,textcomp,microtype}
\usepackage{array}
\usepackage{subfigure}
\usepackage{supertabular}
\usepackage{hhline}
\usepackage[version=3]{mhchem}
\usepackage[detect-all]{siunitx}
\usepackage[pdftex]{graphicx}
\makeatletter
\newcommand\arraybslash{\let\\\@arraycr}
\makeatother
\usepackage{rotating}
\usepackage{placeins}

\usepackage{authblk}

\usepackage{anysize}
\marginsize{1.5cm}{1.5cm}{1.5cm}{1.5cm}

\graphicspath{{figs/}}

\title{The pH sensitivity of solvated tectomer electronics}

\author[1]{Alessandro Chiolerio}
\author[2]{Carsten Jost}
\author[3]{Thomas C.\ Draper}
\author[3]{Andrew Adamatzky}

\affil[1]{Istituto Italiano di Tecnologia, Center for Sustainable Future Technologies, \protect\\  Via Livorno 60, 10144 Torino, Italy}
\affil[2]{PlasmaChem, Schwarzschildstr. 10, D-12489, Berlin, Germany}
\affil[3]{Unconventional Computing Laboratory, University of the West of England, \protect\\  Coldharbour Lane, Bristol, BS16 1QY, UK}

\begin{document}

\maketitle

\begin{abstract}
\noindent Colloidal liquid robotics with embodied intelligence solutions mimicking biologic systems, in response to the future increasingly distributed sensing and the resulting data to be managed, has been proposed as the next cybernetic paradigm. Solutions for data storage and readout \textit{in liquido} require a physical structure able to change configuration under electrical stimuli.
    We propose tectomers as a candidate for such an adaptive structure. A tectomer is a oligomer made of few oligoglycine units with a common centre. Tectomers undergo pH dependent assembly in a single layer supramers. Tectomers represent a stable paradigm, in their amorphous or crystalline forms, reversibly influenced by solution pH, whose electronic properties are studied herein.
\end{abstract}

\begin{quote}
In memory of Alexey Kalachev, founder of PlasmaChem, the only supplier of tectomers.
\end{quote}

\section{Introduction}

Liquid robotics, in response to the future increasingly distributed sensing and the resulting data to be managed, has been proposed as the next cybernetic paradigm~\cite{chiolerioquadrelli2017}. Envisaged solutions require the development of energy management subsystems~\cite{chiolerioquadrelli2018}, mobility subsystems, liquid state sensors, and liquid state logic devices and memories~\cite{chiolerioroppolo2016}, most importantly. 

Several species of liquid robots have been proposed so far. Chemotactic droplet robots move along gradients of Marangoni flow~\cite{vcejkova2017droplets}. Hardware robots can be equipped with liquid brains~\cite{adamatzky2003liquid}, thus only actuators and interface are conventional. Robots based on electro-rheological fluids propulse due to sol-gel phase transitions~\cite{sadeghi2012innovative}. 
Oil micro-droplets move  in a solution of a cationic surfactant~\cite{hirono2018locomotion}. Swimming wet and soft robots are made of ionic polymer metal composites~\cite{stoimenov2009soft,yeom2009biomimetic}. Liquid metal robots  swim due to interaction of Galinstan with aluminium and the resultant imbalance  of  surface  tension induced  by  the   bipolar  electro-chemical  reaction~\cite{zhang2015self,eaker2016liquid}. 
All existing prototypes of liquid robots move either along chemical, thermal, electrostatic, magnetostatic or electro-magnetic gradients or due to geometrical constraining of space. Their environment is pre-programmed. To enable the liquid robots with an ability to make a choice we must equip them with a liquid/colloid brain. Purely liquid controllers, being a subset of liquid computers~\cite{adamatzky2018dry}, can be implemented as (micro-)fluidic logical circuits or reaction-diffusion media. Fluidic devices are realised on elaborate geometrical constraining and reaction-diffusion controllers are slow and have a short lifetime. Another way to implement computing and memory devices in the liquid phase is to select biochemical species which undergo controllable assembly and disassembly. 
Their intrinsic parallelism, due to the presence of a number of elementary processing units (the molecules or monomers dispersed in a solvent), suggest that a liquid state logic device could be used with non-conventional architectures, such as Holonomic ones~\cite{pribram2013brain}. Solutions for data storage and readout \textit{in liquido} require a physical support able to change configuration under electrical stimuli. We propose that data storage can be implemented with tectomers.

\begin{figure}[htb]
    \centering
    \subfigure[]{\includegraphics[scale=0.8]{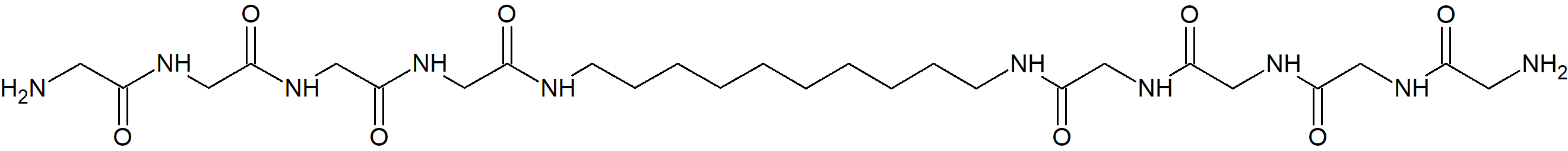}}\\
    \subfigure[]{\includegraphics[scale=0.8]{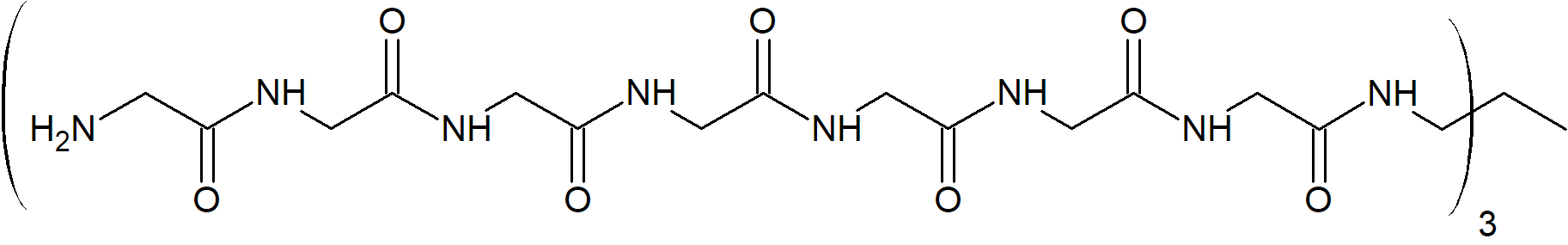}}\\
    \subfigure[]{\includegraphics[scale=0.8]{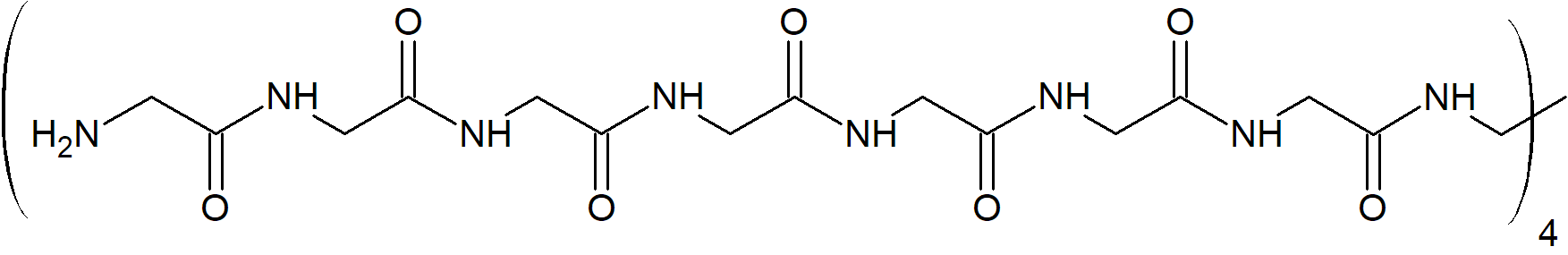}}\\
    \subfigure[]{\includegraphics[width=0.2\textwidth]{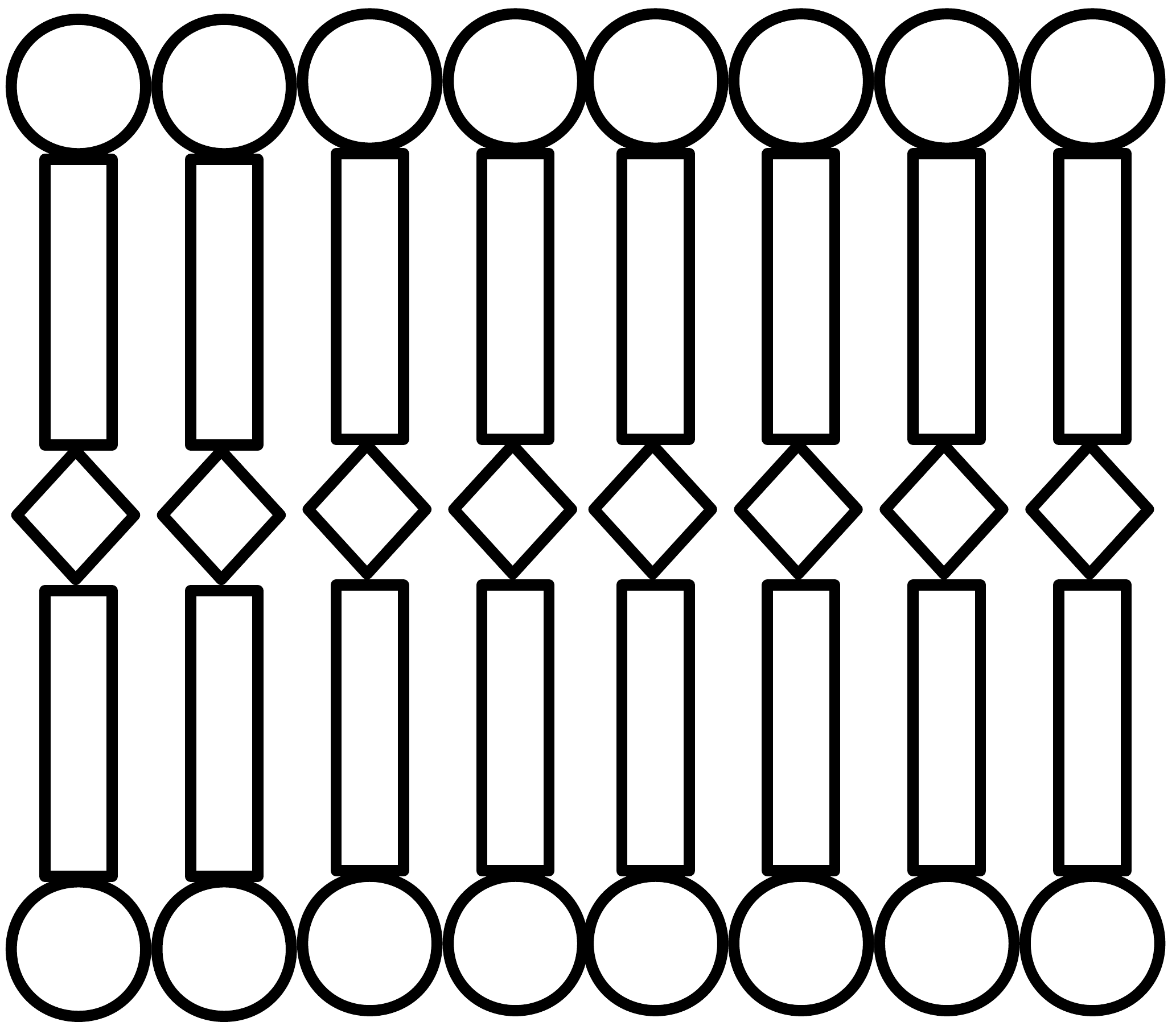}}
    \caption{Tectomers. Chemical structure of (a)~2-tailed (T2), (b)~3-tailed (T3) and (c)~4-tailed (T4) tectomers. (d)~Supramer. $\diamond$ is \ce{-(CH_2)_n-}, rectangle is olygoglycine tail, and $\circ$ is an amino group. The scheme is redrawn from \cite{tsygankova2014biantennary}.}
    \label{fig:tectomersStructure}
\end{figure}

A tectomer is an olygoglycine molecule comprised of oligoglycine units linked to one common centre (Fig.~\ref{fig:tectomersStructure}abc)~\cite{tuzikov2003polyglycine,gorokhova2006spontaneous}.
The oligoglycines form polyglycine II-type structures, stabilised by hydrogen bonds~\cite{bovin2008}. Tectomers assemble into stable and rigid regular layers of mono-molecular thickness (Fig.~\ref{fig:tectomersStructure}d)~\cite{tuzikov2003polyglycine,tsygankova2014biantennary}.  As demonstrated in~\cite{gyurova2017self}, the  positively charged structures with  high electrophoretic mobility are formed  up to 30th minute after preparation of the solutions.
The sizes of tectomers and their electrical charges remain preserved for at least 24  hours. 
Two-dimensional platelets of tectomers have been previously functionalised with one-dimensional nanowire network of silver nanowires and transparent electrodes have been obtained~\cite{jurewicz2018functionalization}. The tectomers brought several advantages: increased conductivity of the network by strengthening contacts between nanowires, protected silver nanowires against oxidation and sulfurisation, prevented detachment of nanowires in case of immersion in aqueous medium~\cite{jurewicz2018functionalization}.

The assembly of tectomers is pH dependent. Increase of pH leads to aggregation and assembly into ordered (crystalline) structures; while decrease of the pH causes dissolution and amorphous structures~\cite{garriga2016two}. 
Therefore the stable aggregation state and the self-assembly capability of tectomers could be exploited to realise a pH-stabilised system with electrical read-out. In the following section~\ref{Methods}, details about our experimental setup are given; while in section~\ref{Results}, details about electronic properties characterisation are presented.

\section{Methods \& Materials}
\label{Methods}

Lyophilised 2-, 3- and 4-tailed tectomers were supplied by PlasmaChem (Berlin, Germany), potassium phosphate monobasic was provided by Sigma Aldrich, and sodium hydroxide was provided by Fisher Scientific. All chemicals were used without further purification. Throughout this text lyophilised 2-, 3- and 4-tailed tectomers are denoted as T2, T3 and T4 respectively. To prepare a solution of tectomers, \SI{1.0}{\milli\gram} of the desired tectomer was dissolved in \SI{1.00}{\milli\L} of either deionised water (DIW, \SI{15}{\mega\ohm\centi\metre}) or a buffered aqueous solution. To aid dissolution, samples were subjected to an ultrasonic bath (Ultrawave U300H, \SI{35}{\watt}, \SI{30}{\kilo\hertz}) for 15 minutes. Aqueous buffer solutions with a pH of 5.9 and 8.2 were both prepared using DIW, \ce{KH2PO4} (0.10~M) and adjusted using \ce{NaOH}. The pH was measured using a calibrated Perkin Elmer P200-02 pH Meter. All experiments were performed using \SI{10.0}{\micro\L} droplets.

Electrical stimulation and recording performed via needle electrodes with twisted cables were used.\footnote{SPES MEDICA SRL, Via Buccari 21, 16153, Genova, Italy} These electrodes were inserted \SI{1}{\milli\metre} deep into the tectomer solution droplet, and the distance between them kept at \SI{2}{\milli\metre}.  Fresh electrodes were used between each droplet, to prevent the effects of droplet evaporation and tectomer coating of the electrodes. The electronic characterization was assessed using a SCS4200 from Keithley with triaxial cables and preamplifiers. DC characterisation was done in the range up to [-1.2, +1.2]~V. This limit was chosen to avoid the electrolysis of water at \SI{1.23}{\V}. AC measurements have been taken in the range from 1~kHz to 10~MHz, typically with a signal amplitude of 10~mV RMS. By adding a DC bias and sweeping it between two saturation values (-0.5~V and +0.5~V), impedance is measured at a fixed frequency with a small signal over-imposed (10~mV RMS). PT (Pulse Train) measures feature a high speed pulse generator with internal reference and feedback system that is capable of measuring its own output, plus two independent high speed voltage and current units connected to the DUT. By simultaneously monitoring the voltage and current at each of the electrodes, the real resistive response is measured under the application of multiple unipolar pulses. The pulse is rectangular, has an amplitude of 0.5~V and a duration of 1~ms, selected to be comparable to biologic action potentials. The electrical response is recorded for further 200~$\mu$s, enabling us to trace the decay of the sample to rest (0~V) condition. During each measure the sample is submitted to a train of 50 pulses separated by 4~ms idle time and the resistive response is recorded and internally averaged; this procedure is repeated 10 times to trace any eventual drift and averaged to compute standard deviation.

\section{Results}
\label{Results}

\subsection{DC measurements}


\begin{figure}[htpb]
    \centering
    \subfigure[]{\includegraphics[scale=0.075]{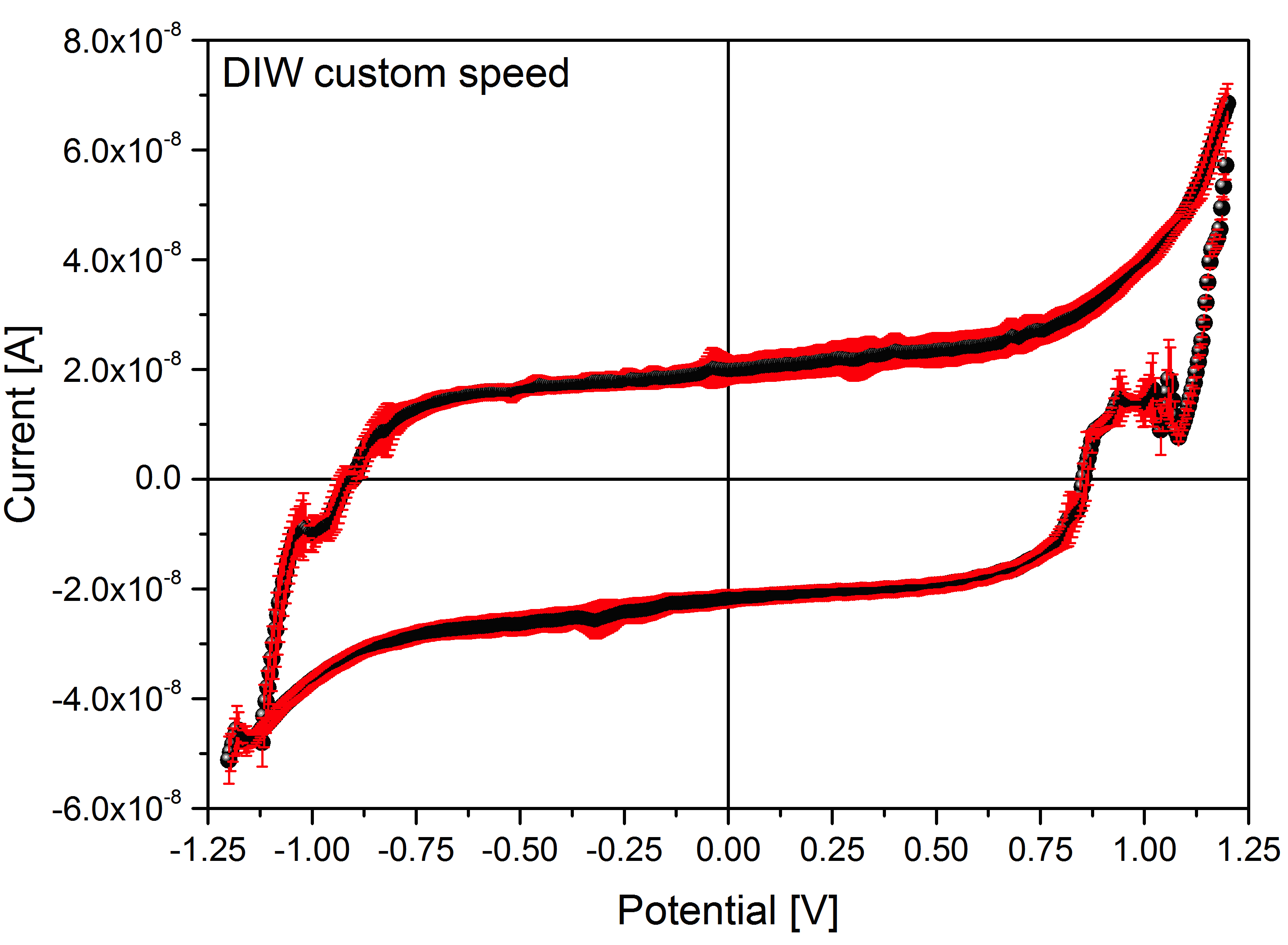}\label{fig1}}
    \subfigure[]{\includegraphics[scale=0.075]{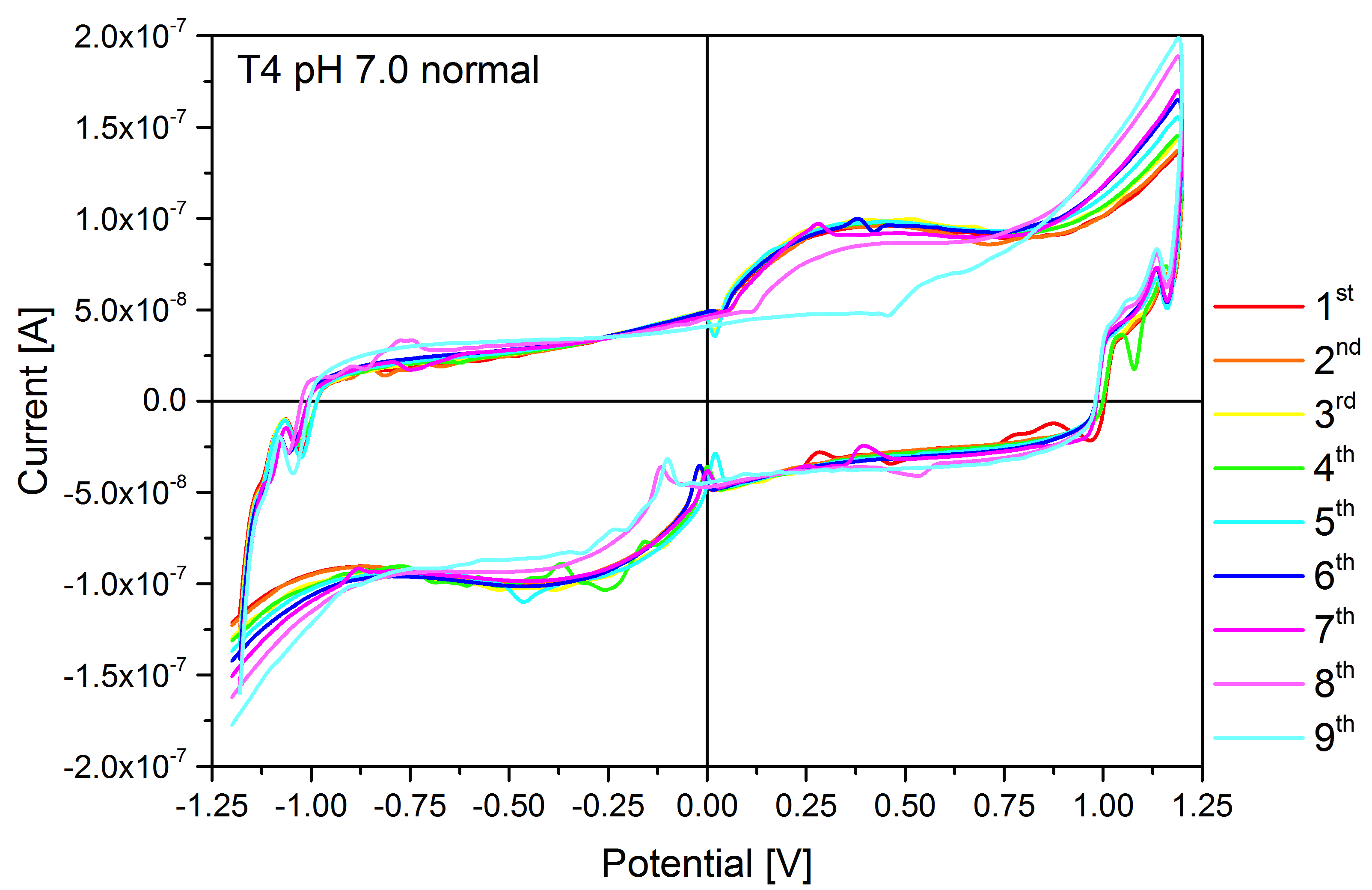}\label{fig2}}
    \caption{IV cycle: 
    (a)~Pure DIW average IV curve.
    (b)~T4 pH=7.0, nine cycles of IV.
    }
    \label{fig:12}
\end{figure}

DC properties of the different tectomers in DIW solutions at different pH have been assessed. Figure~\ref{fig1} shows the average IV cycle measured by testing pure DIW, in an extended range of 1.2~V. No higher potentials were produced to avoid water decomposition and related electrical instability. The cycle shows a relevant hysteresis, due to the droplet associated parasitic capacitance, and a maximum current of about 70~nA flowing through the sample.

For comparison, Fig~\ref{fig2} shows the response of a T4 sample at neutral pH, the first nine cycles are plotted. During measurements the sample re-configures the order of molecules, so that the maximum current flowing is increased from about 120 of the first cycle to about 180~nA of the last. Furthermore the shape of the curve, though retaining qualitatively the features seen with DIW, reveals other phenomena.

\begin{figure}[htpb]
    \centering
    \subfigure[]{\includegraphics[scale=0.074]{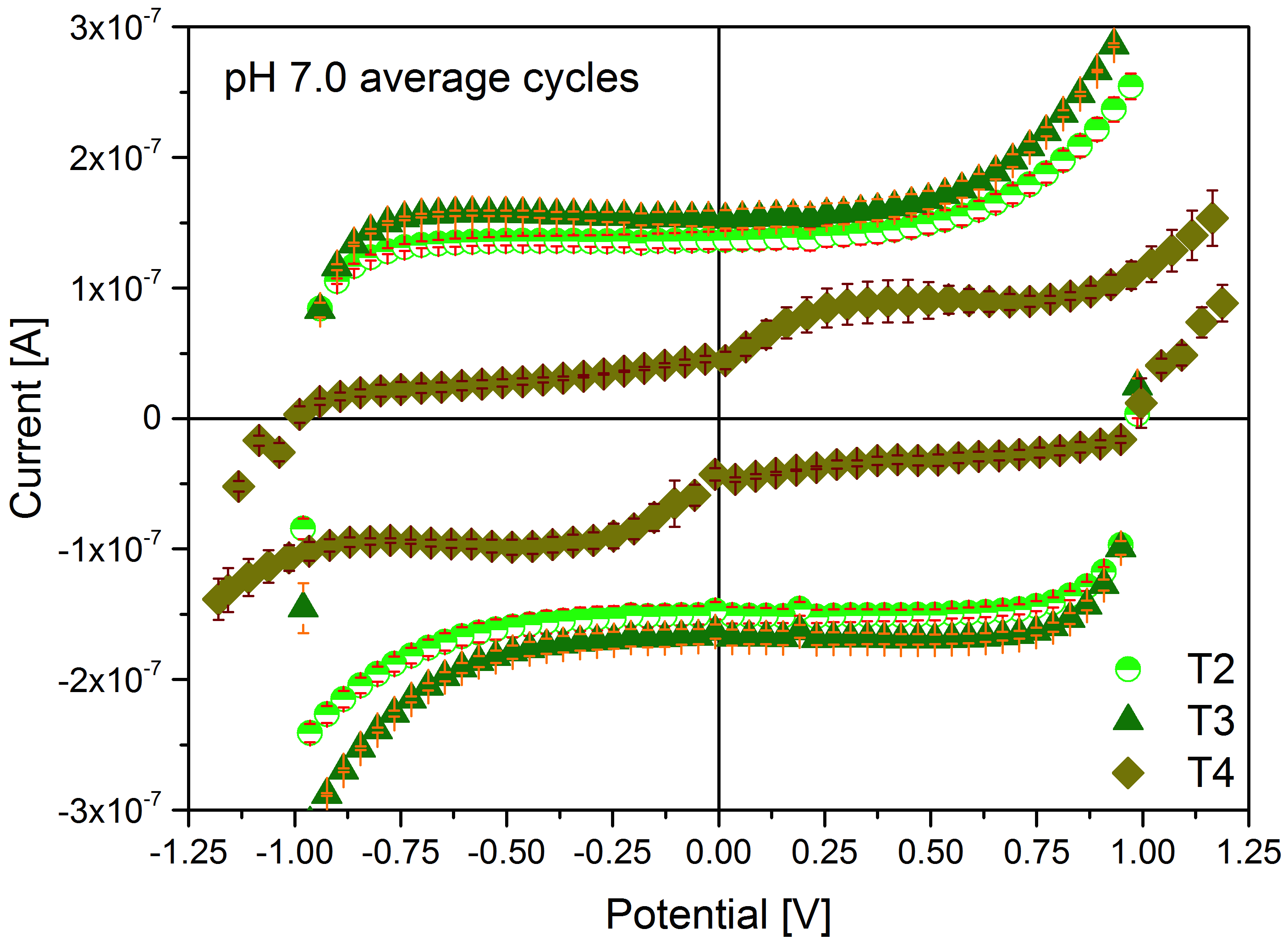}\label{fig3}}
    \subfigure[]{\includegraphics[scale=0.074]{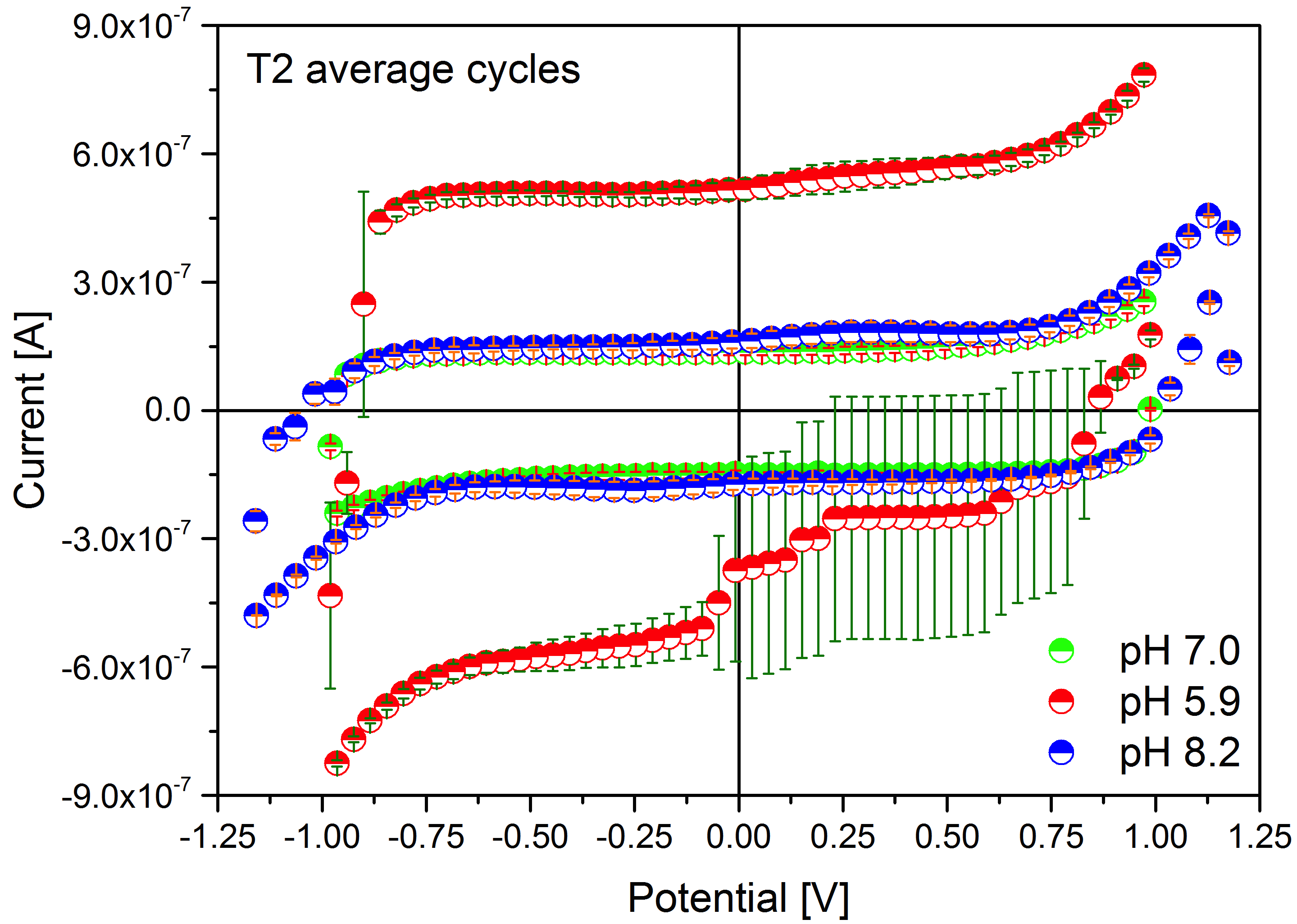}\label{fig4}}
    \subfigure[]{\includegraphics[scale=0.074]{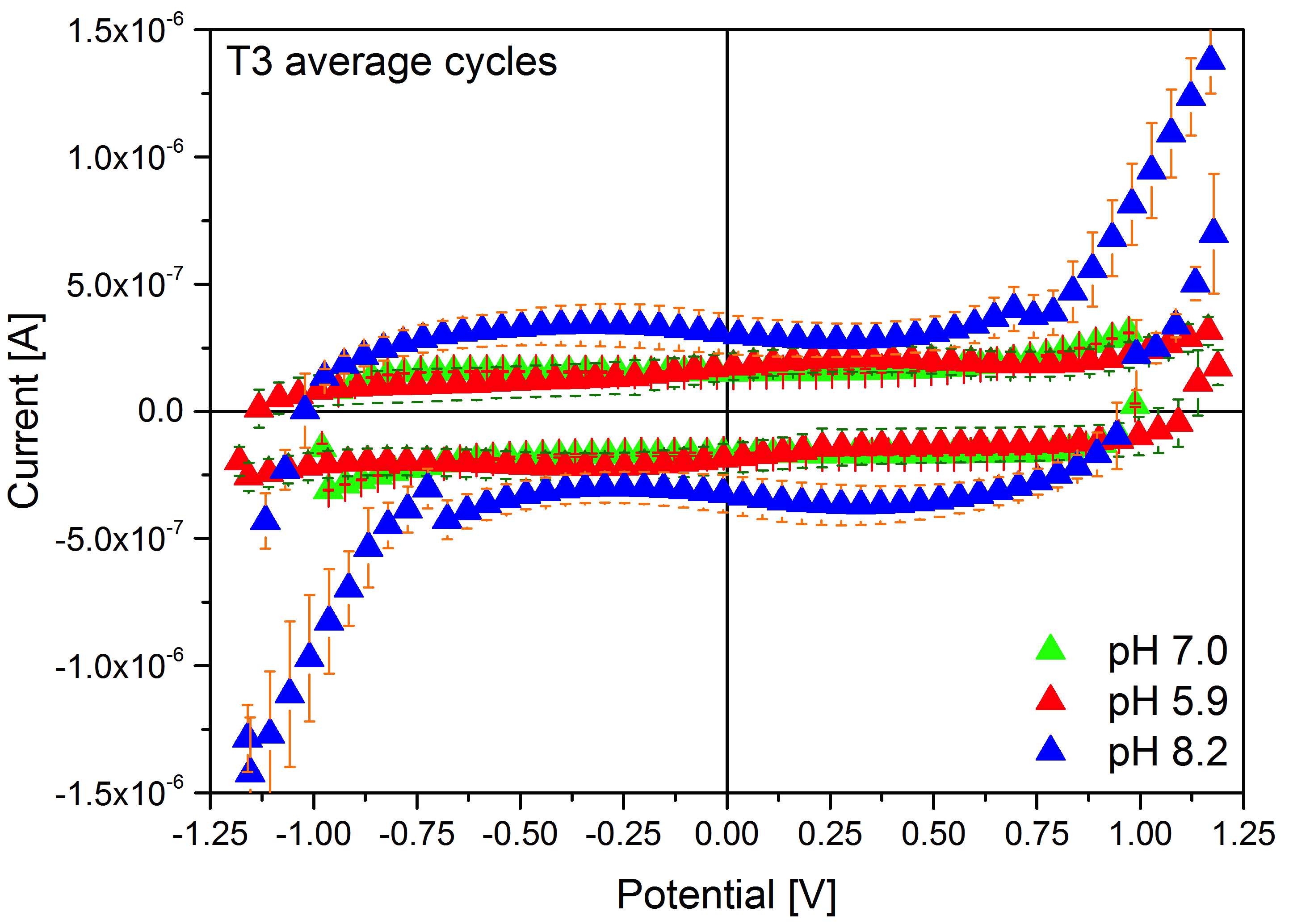}\label{fig5}}
    \subfigure[]{\includegraphics[scale=0.074]{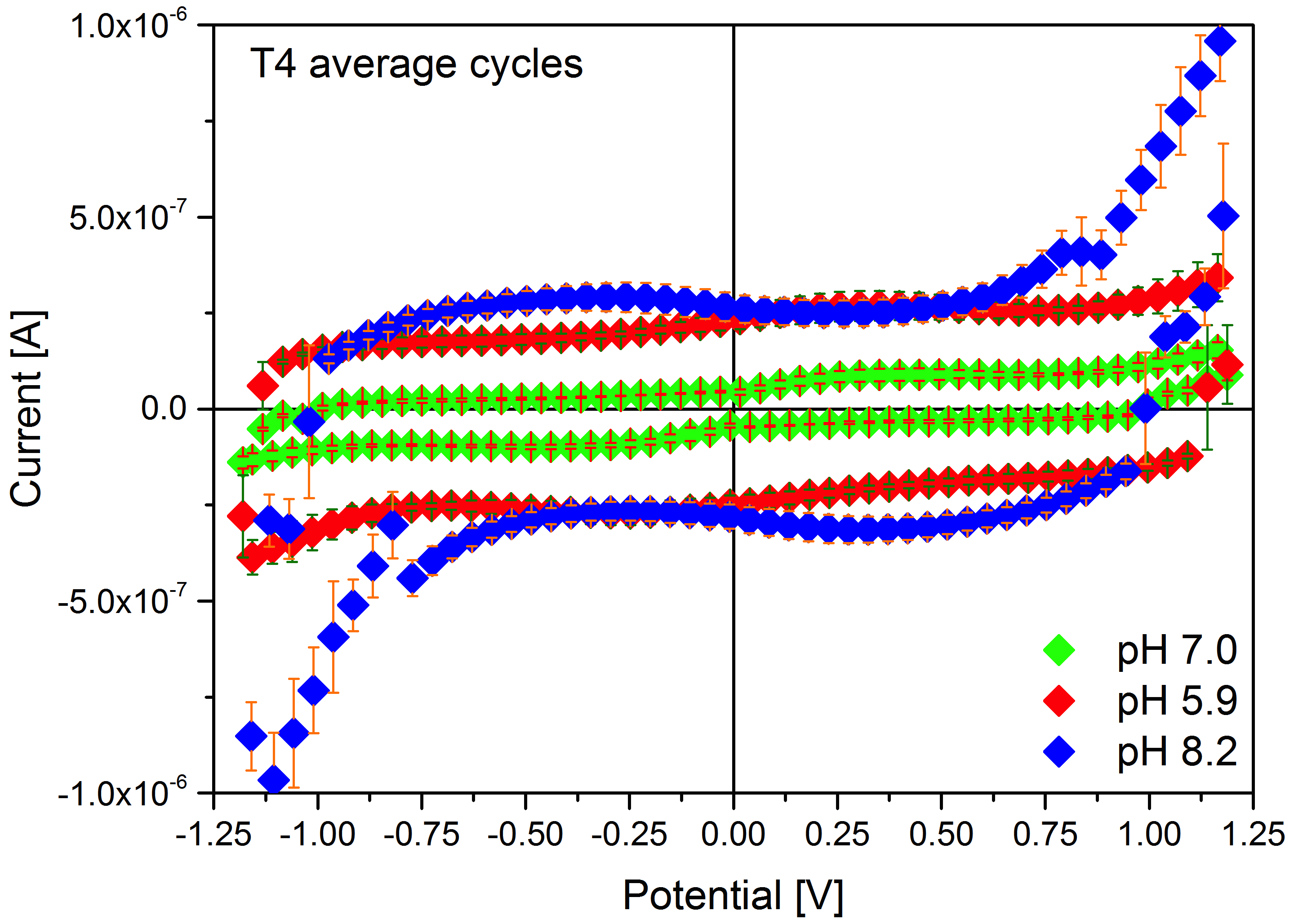}\label{fig6}}
    \caption{
    (a)~Comparison between T2, T3 and T4 at pH=7.0. Only 1 point every 10 is shown, for clarity.
    (b)~T2, comparison between pH=5.9, 7.0 and 8.2. Only 1 point every 10 is shown, for clarity.
    (c)~T3, comparison between pH=5.9, 7.0 and 8.2. Only 1 point every 10 is shown, for clarity.
    (d)~T4, comparison between pH=5.9, 7.0 and 8.2. Only 1 point every 10 is shown, for clarity.
    }
    \label{fig:134}
\end{figure}

The three tectomers have rather different responses. In Fig.~\ref{fig3} their average IV curves have been compared, under pH=7.0. The cycles opens more for T2 and T3, while T4 features points with negative differential resistance both in quadrant I and III as well as second order discontinuities across the zero potential point. The standard deviation is very low for all the three samples, confirming their stability. Maximum current is about 300~nA for T3 sample. Figure~\ref{fig4} shows a comparison between responses of T2 tectomers in three different pH conditions, basic (pH 8.2), neutral (7.0 pH) and acid (pH 5.9). While the basic and neutral solutions have very similar responses, the acidic condition produces a much more conductive state (approximately three times the others) with an unstable region in quadrant IV where a resistive switching occurs, with a set voltage found in the range (-40~mV, +880~mV). Figure~\ref{fig5} shows the same comparison as above for the T3 molecule. This suggests the application as resistive switching device (RSD), having a unipolar switching (switching occurs only in one portion of the hysteresis loop, here for positive potentials). In this case, the acid and neutral conditions have a very similar response, while the basic solution achieves much higher currents (up to 1.5~$\mu$A) and features in every quadrant negative differential resistance. Figure~\ref{fig6} shows the comparison for the T4 molecule, similar to the case of T3. Here the basic solution has negative differential resistance in quadrant IV.

\subsection{AC measurements}

\begin{figure}[!htpb]
    \centering
    \subfigure[]{\includegraphics[scale=0.075]{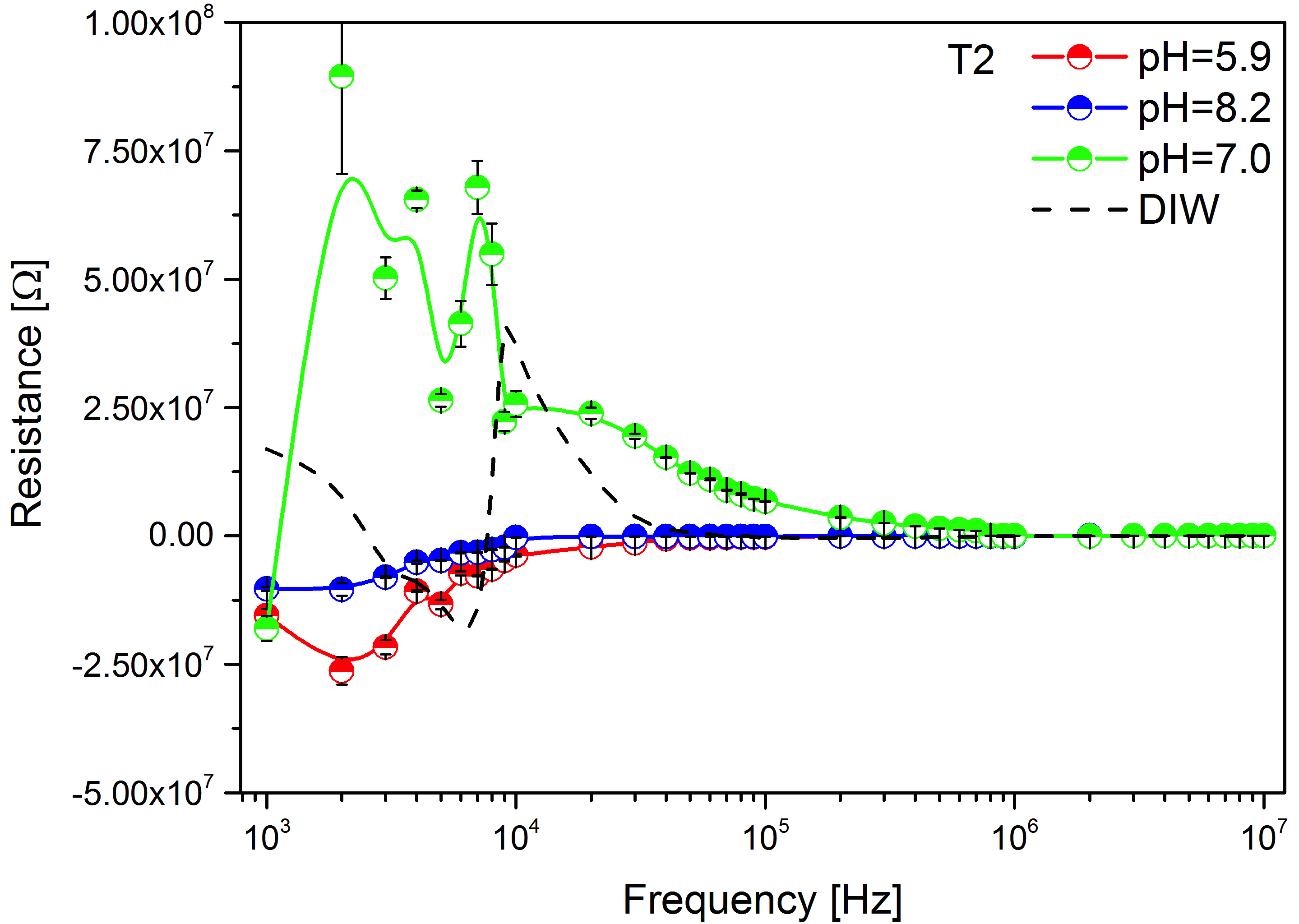}\label{fig7}}
    \subfigure[]{\includegraphics[scale=0.075]{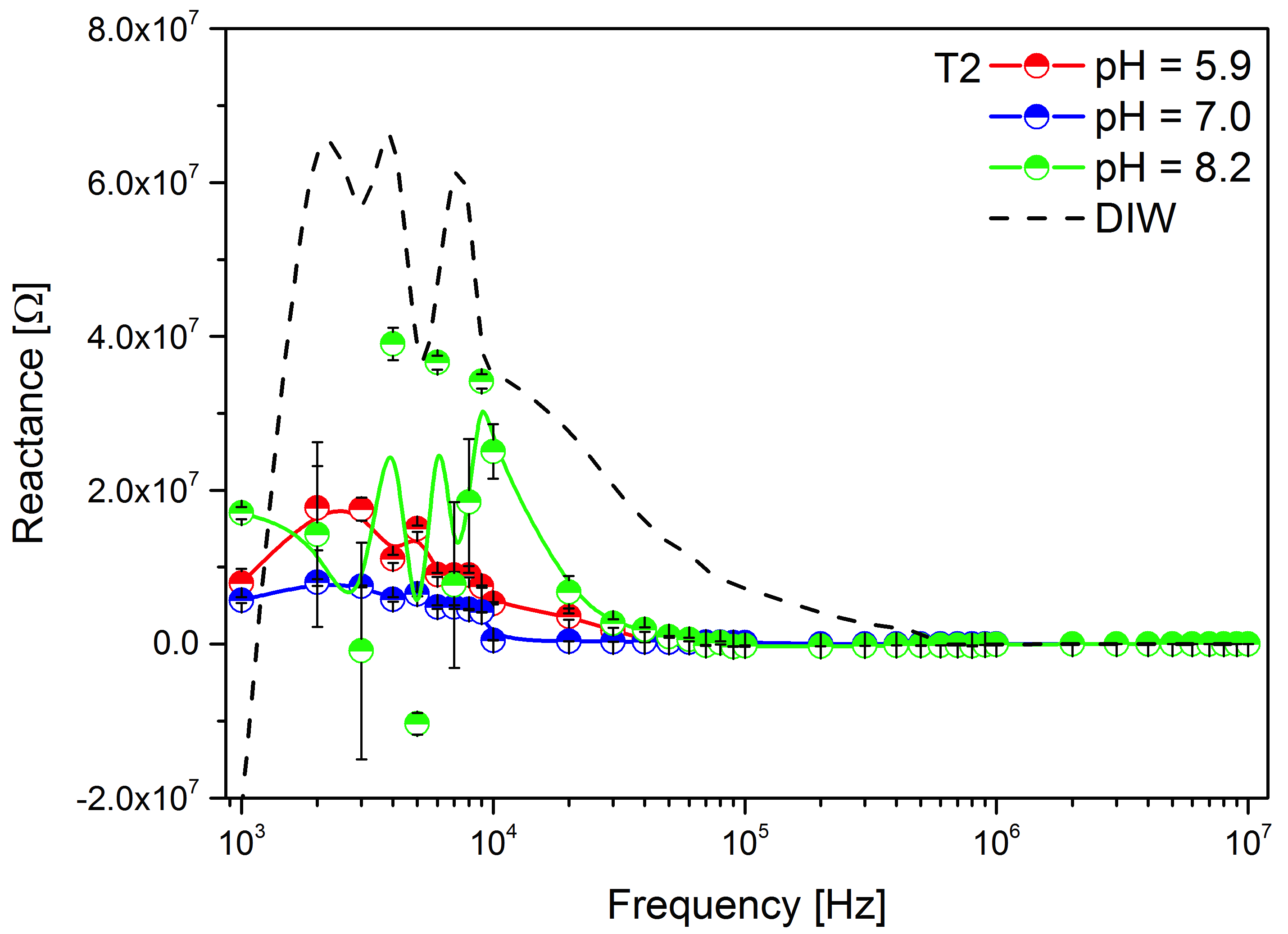}\label{fig8}}
    \subfigure[]{\includegraphics[scale=0.075]{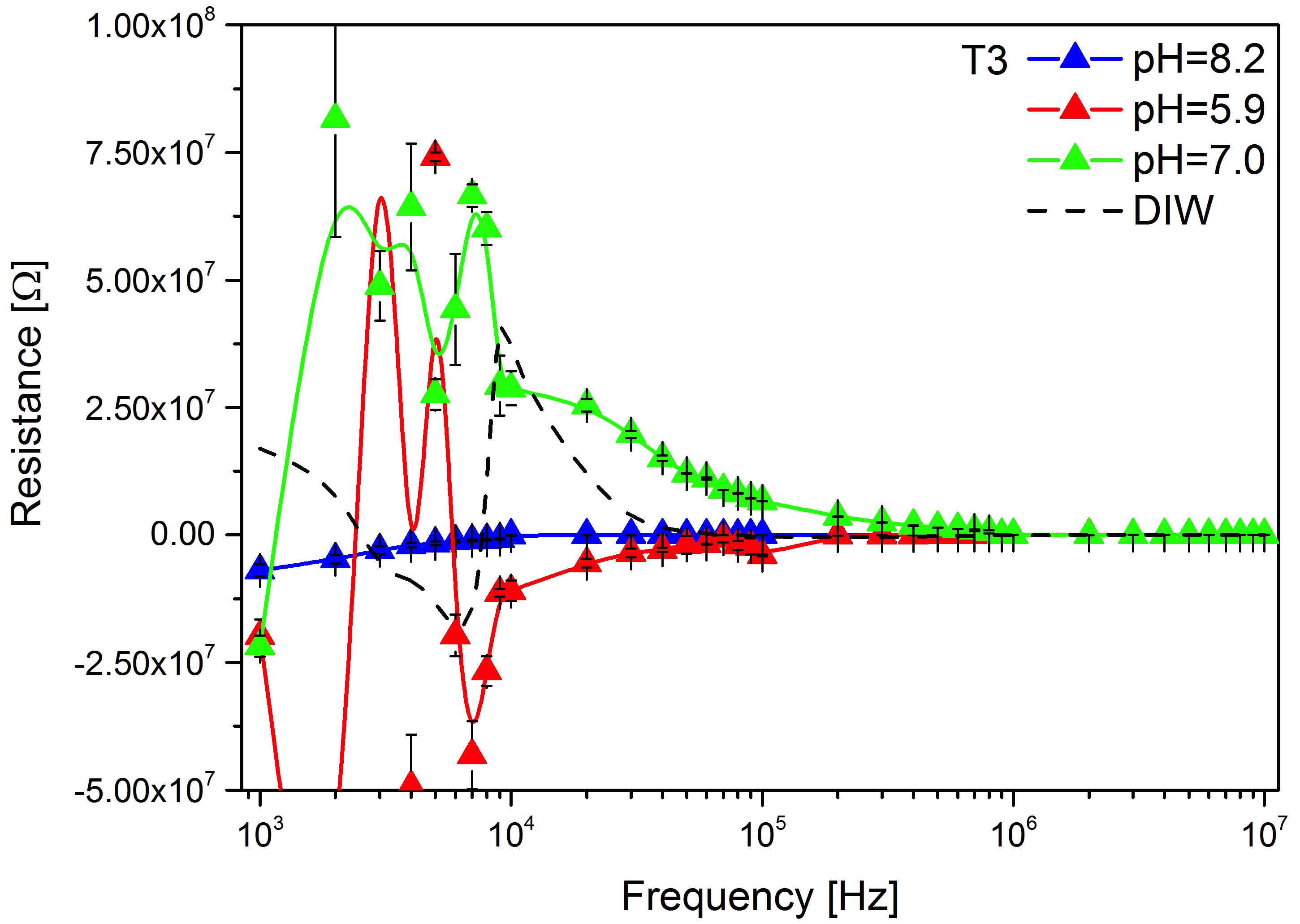}\label{fig9}}
    \subfigure[]{\includegraphics[scale=0.075]{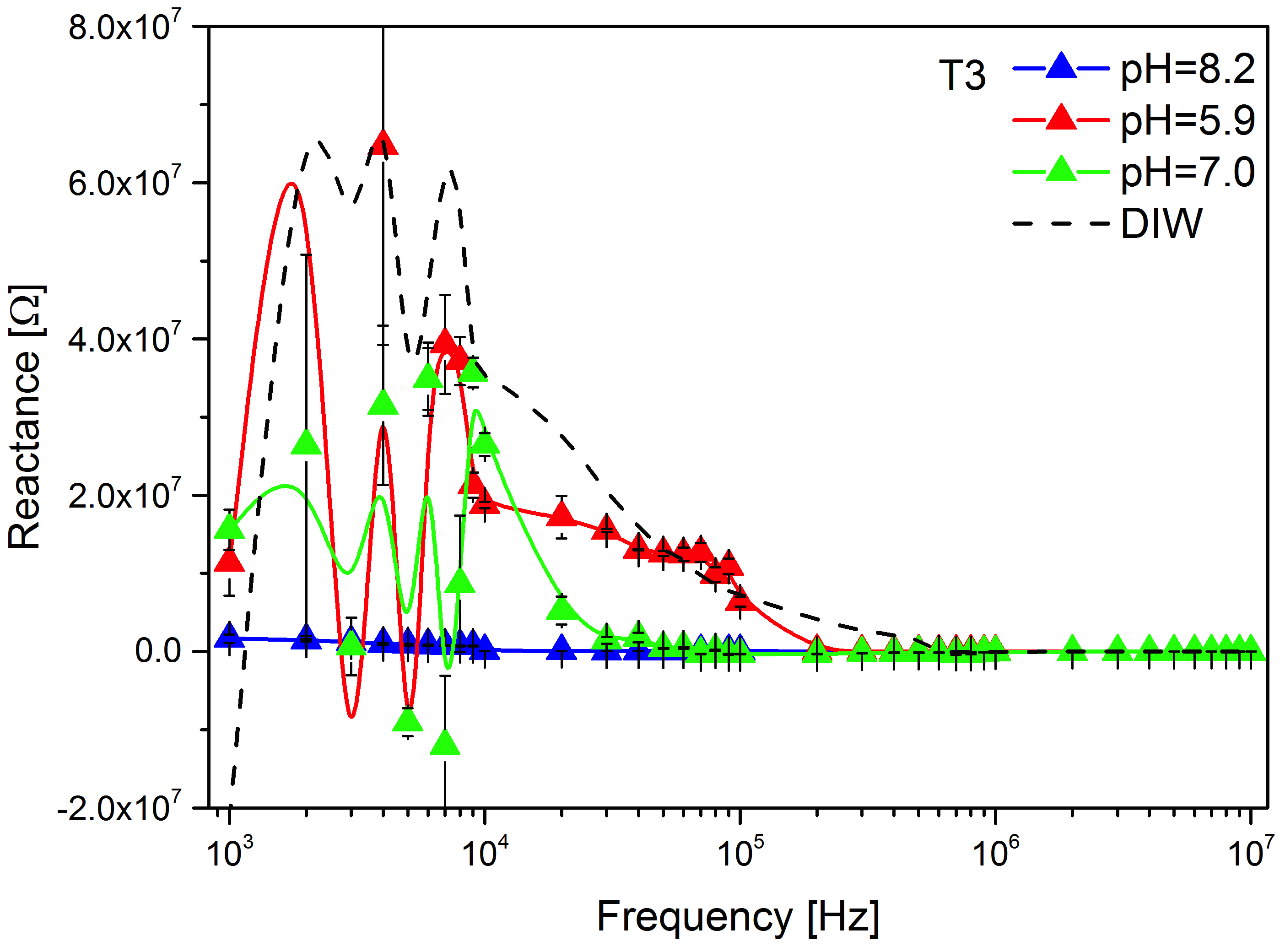}\label{fig10}}
    \subfigure[]{\includegraphics[scale=0.075]{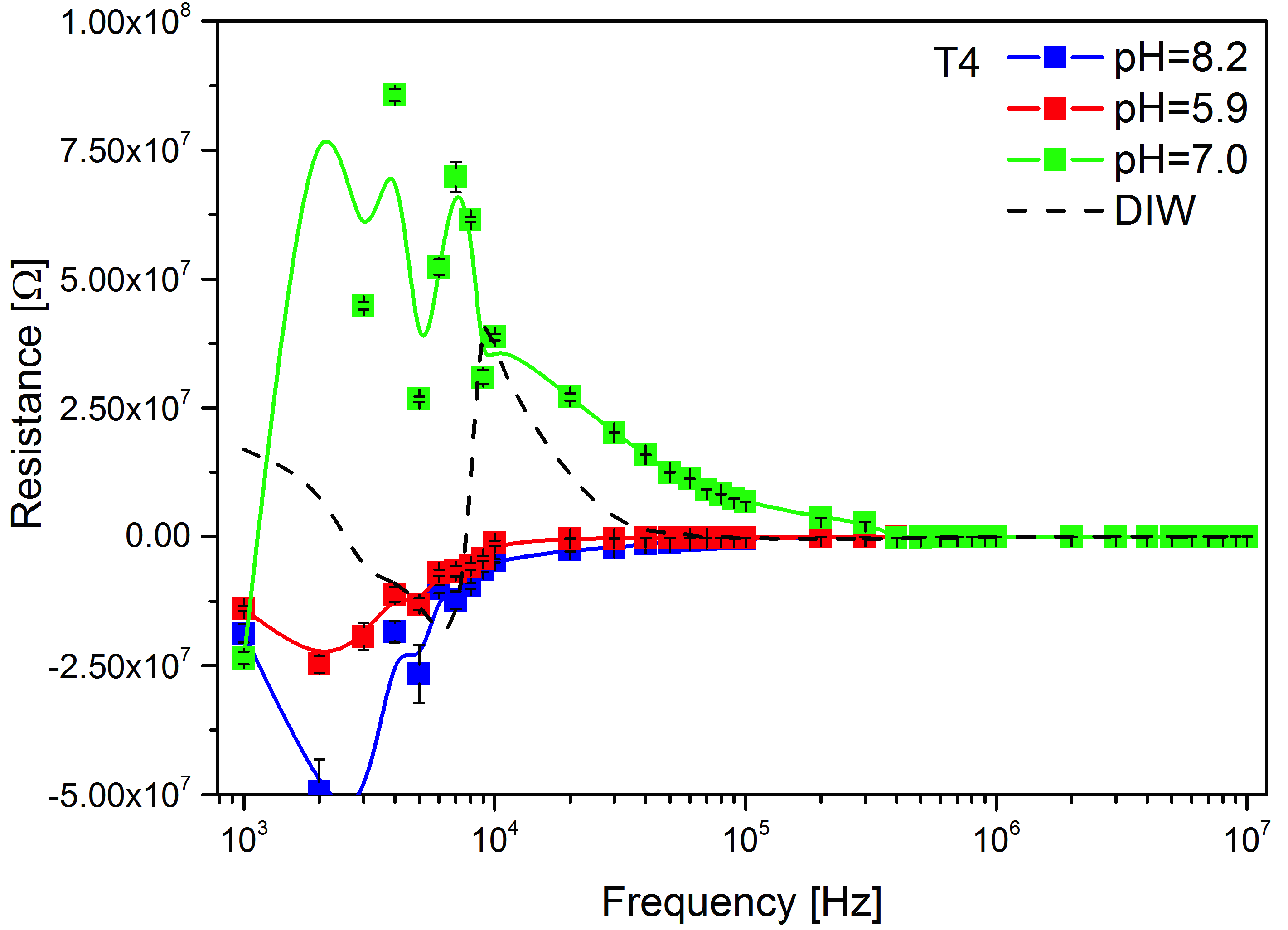}\label{fig11}}
    \subfigure[]{\includegraphics[scale=0.075]{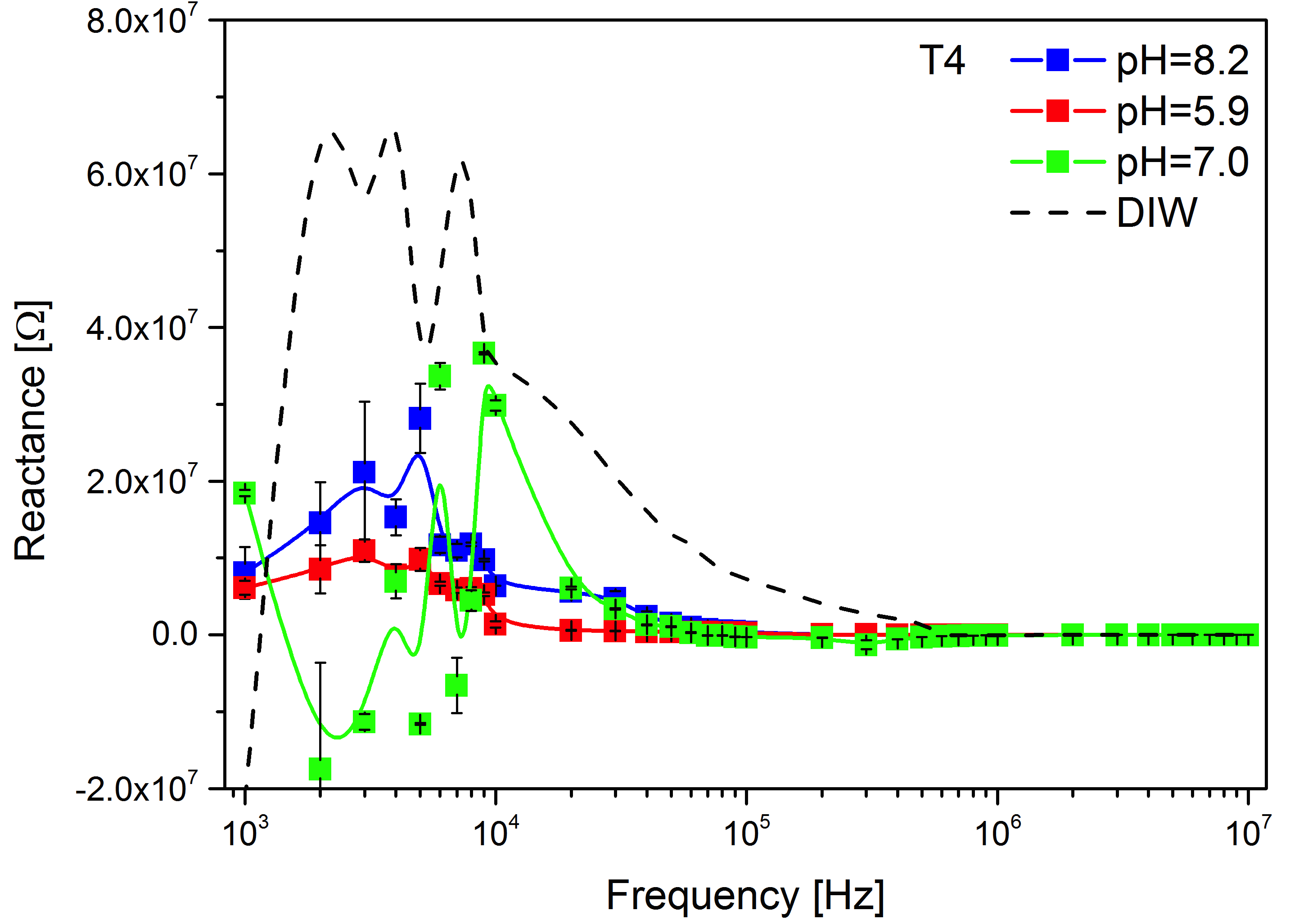}\label{fig12}}
    \caption{Impedance measurements in the range between 1 kHz and 10 MHz, for all tectomer sample T2 (ab), T3 (cd) and T4 (ef) in all pH states and for DIW. Average curves out of multiple measurements, standard deviation also shown.
    }
    \label{fig:7_12}
\end{figure}

Due to connector parasitic impedance it is difficult to measure high impedance materials, therefore as a reference in every plot the curve of DIW has been shown for comparison (Fig.~\ref{fig:7_12}), to help reveal that the measurements are significant. Each curve reports also standard deviation, indicating that the noise level is higher in the range between 1~and~10~kHz. Typically the samples kept at neutral pH show a higher resistance and a very similar response, while both the samples kept in basic and acid pH show a negative resistance range up to 100 kHz as well as some differences between T2, T3 and T4. The reactance of neutral samples shows the fingerprints of peaks found in DIW response (between 1~and~10~kHz). Typically reactance is higher for neutral samples and the smallest for the basic sample; it decreases with frequency, as for predominantly capacitive materials. The only sample with a clear inductive behaviour is T4 in the neutral state, showing  a negative reactance in the low frequency region of the spectrum. Generally speaking, our measures demonstrate that it is possible to distinguish between the different tectomers by measuring their impedance in the range between 1~kHz and 1~MHz.

\subsubsection{Frequency response}

\begin{figure}[!htpb]
    \centering
    \subfigure[]{\includegraphics[scale=0.075]{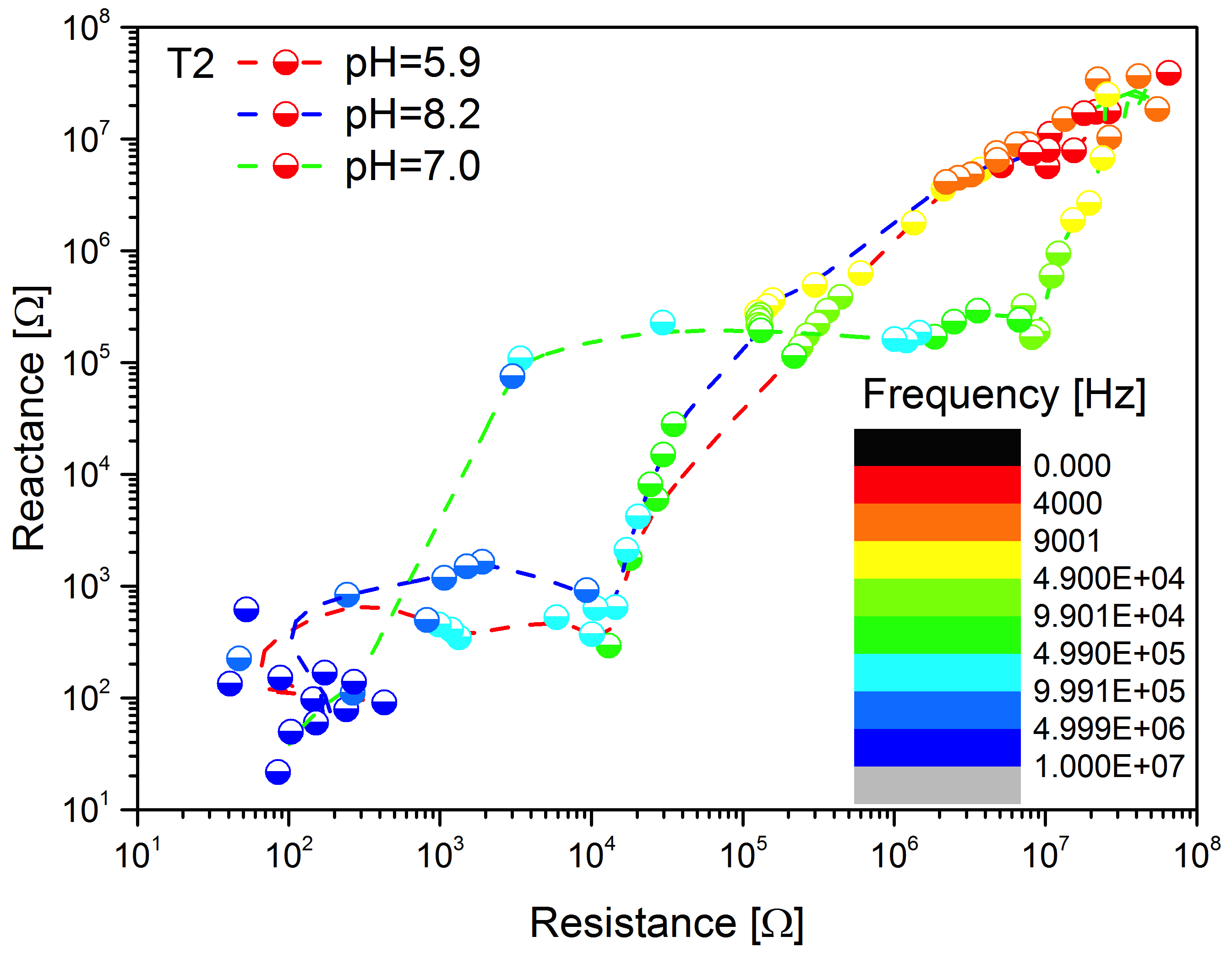}\label{fig13}}
    \subfigure[]{\includegraphics[scale=0.075]{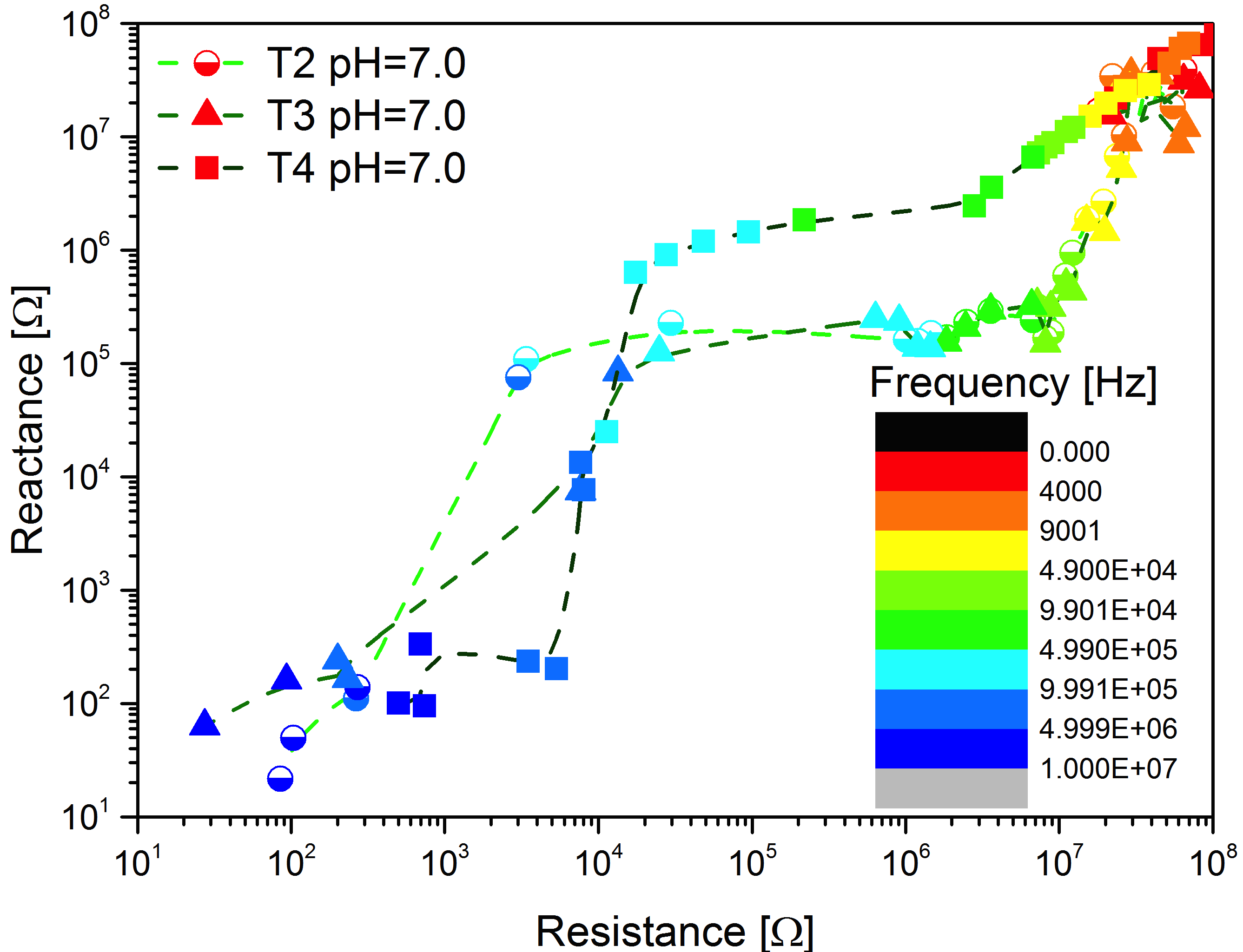}\label{fig14}}
    \caption{Nyquist plot for (a)~sample T2 in three pH conditions and (b)~ samples T2, T3 and T4 in the neutral state. Both resistance and reactance shown in logarithmic scale after taking absolute value.
    }
    \label{fig:13_14}
\end{figure}

As alternative view, the Nyquist plot includes both the real and imaginary components of impedance (absolute value) on the same plane and shows frequency as a colour scale. In Fig.~\ref{fig13} the impedance of sample T2 is shown under the three pH values; increasing the signal frequency, reading the plot from top right to low left, both real and imaginary components are reduced. The non-neutral solutions retain a lower impedance with respect to the neutral one in the low frequency range of spectrum, until 50~kHz, in reason of their higher ion contribution to conductivity. Interestingly the reactance of the neutral sample is almost constant from 50~kHz up to about 1~MHz, making it very easy to discriminate between neutral and non-neutral conditions. 
Figure~\ref{fig14} compares the Nyquist plots of the three tectomers in their neutral state. T2 and T3 show a very similar response up to 1 MHz, including the reactance stability plateau. T4 has a similar curve, the reduction of impedance vs frequency featuring a more gentle slope and a stability plateau, in a similar frequency range, placed at higher reactance values. Therefore the range between 100~kHz and 1~MHz could be used to discriminate between different tectomers.

\begin{figure}[!htpb]
    \centering
    \subfigure[]{\includegraphics[scale=0.075]{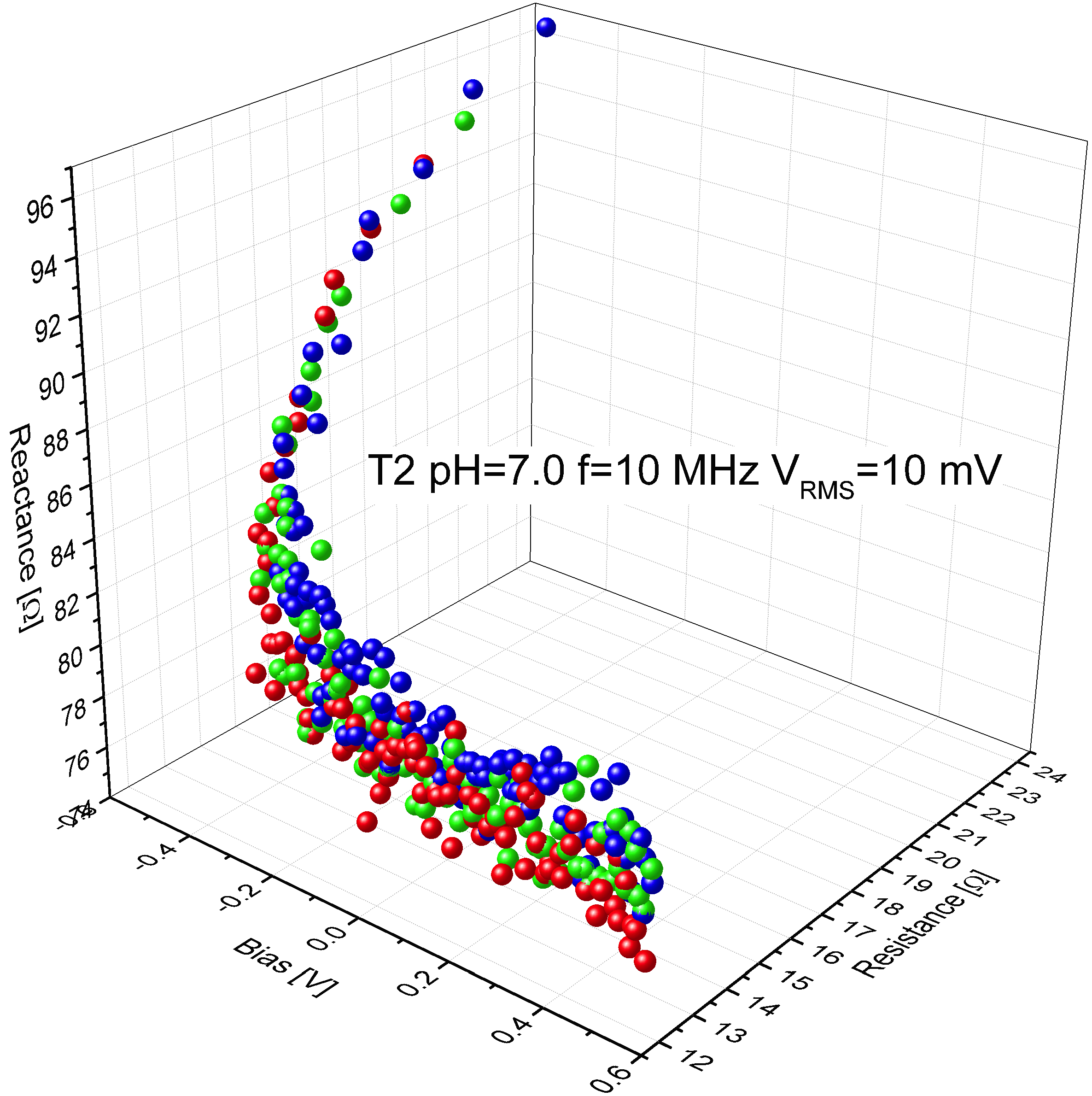}\label{fig15}}
    \subfigure[]{\includegraphics[scale=0.18]{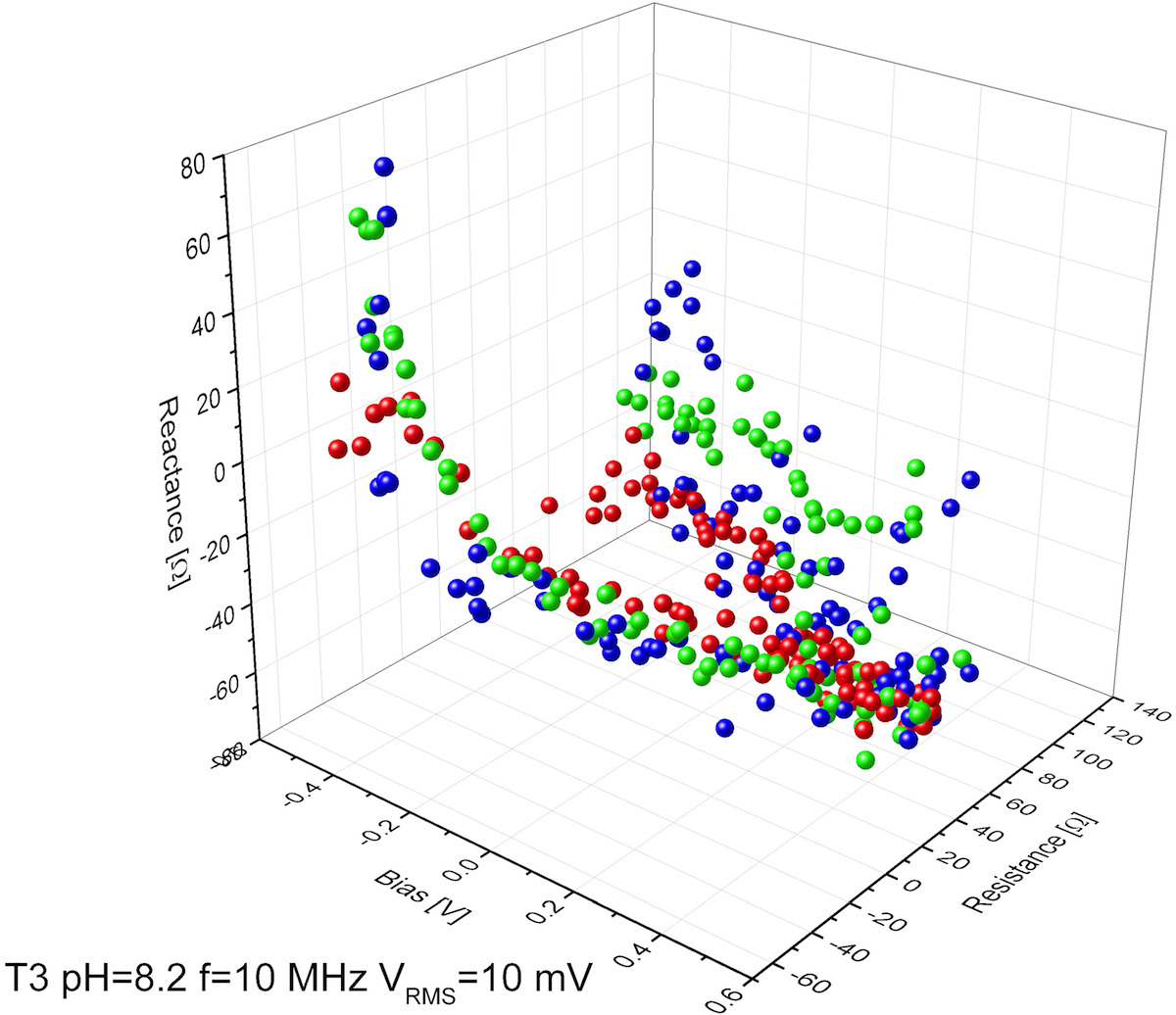}\label{fig16}}
    \subfigure[]{\includegraphics[scale=0.18]{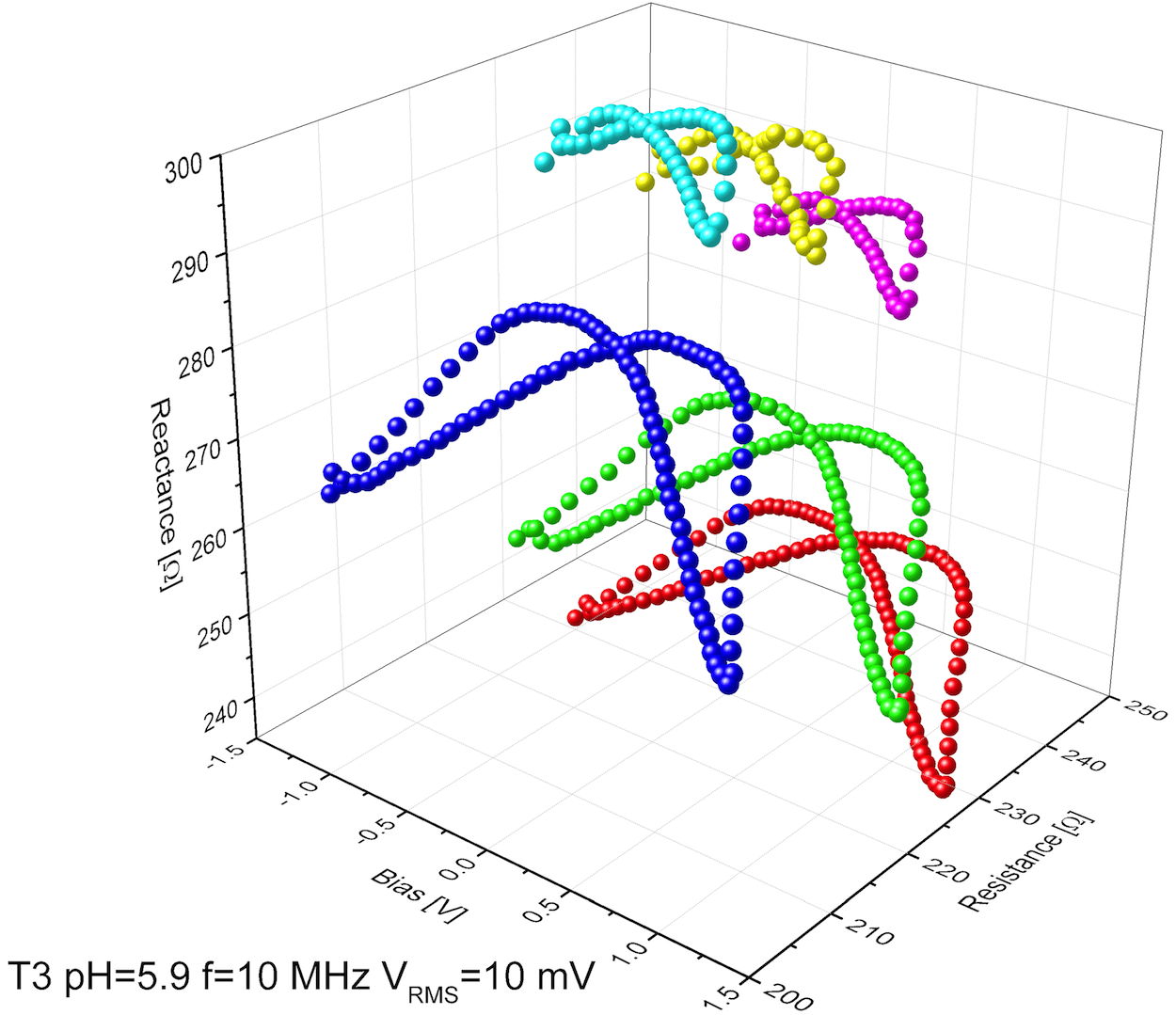}\label{fig17}}
    \caption{3D plots showing resistance and reactance versus bias at 10~MHz and fixed signal amplitude (10~mV), for sample T2 pH~7.0 (a), T3 pH~8.2 (b) and T3 pH~5.9 (c). Colour code identifying measures on different droplet samples.
    }
    \label{fig:15_17}
\end{figure}

\subsubsection{Bias response}

 Building on top of AC stimulus a DC bias,  no variation is seen at 1~kHz and 100~kHz (impedance is not a function of DC bias), while interesting features appear when samples are characterised at 10~MHz. This applies to all the pH conditions, producing patterns of different shape. Results are conveniently shown in Fig.~\ref{fig:15_17}, a 3D graph where DC bias, resistance and reactance represent the three axes (fixed frequency and fixed signal amplitude), while the colour code identifies different measurement sets. Taking T2 as example, the bias acts on impedance by reducing both resistance and reactance, moving from negative to positive potentials. More interesting results are found in the basic form of T3, where a hysteretic behaviour is seen, resulting in two branches that during the sweep have similar reactance but resistance of opposite sign, opening potential applications as voltage controlled gain passive device. The most peculiar behaviour is found in the acid condition for T3, where a butterfly chaotic response is recorded. Such butterfly diagrams in impedance are associated with ferroelectric response~\cite{islam2011}, therefore this is the fingerprint of electrical polarisability of tectomer molecules excited by the signal oscillation at 10 MHz. Some cycles acquired in the 0.5 V range have shown a stable slightly asymmetric butterfly response, at different impedance values (chaotic), while other cycles acquired in the 1.2 V extended range have shown a stable chaotic symmetric butterfly response.

\subsection{Pulse Train (PT) measurements}

\begin{figure}[!htpb]
    \centering
    \subfigure[]{\includegraphics[scale=0.075]{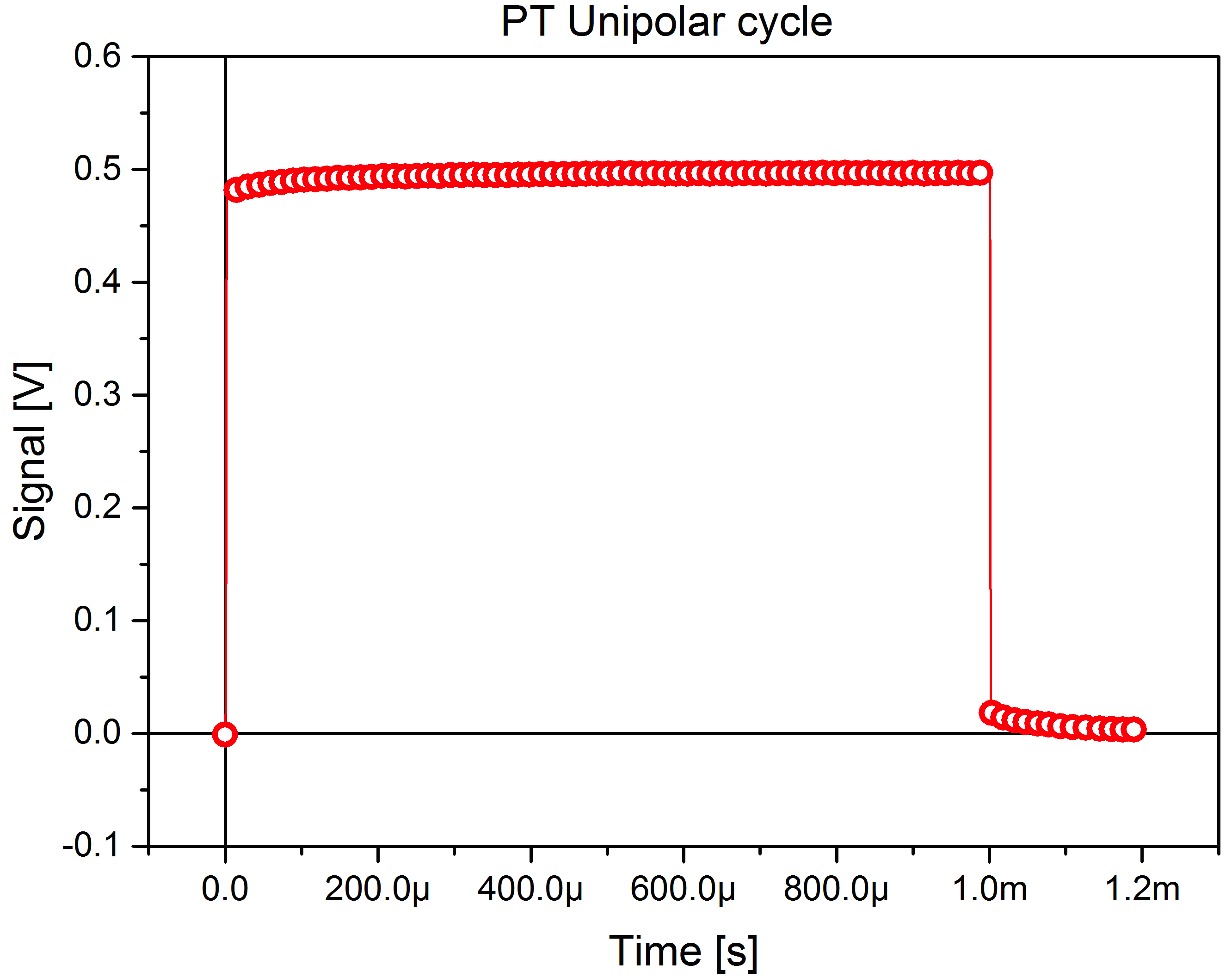}\label{fig18}}
    \subfigure[]{\includegraphics[scale=0.075]{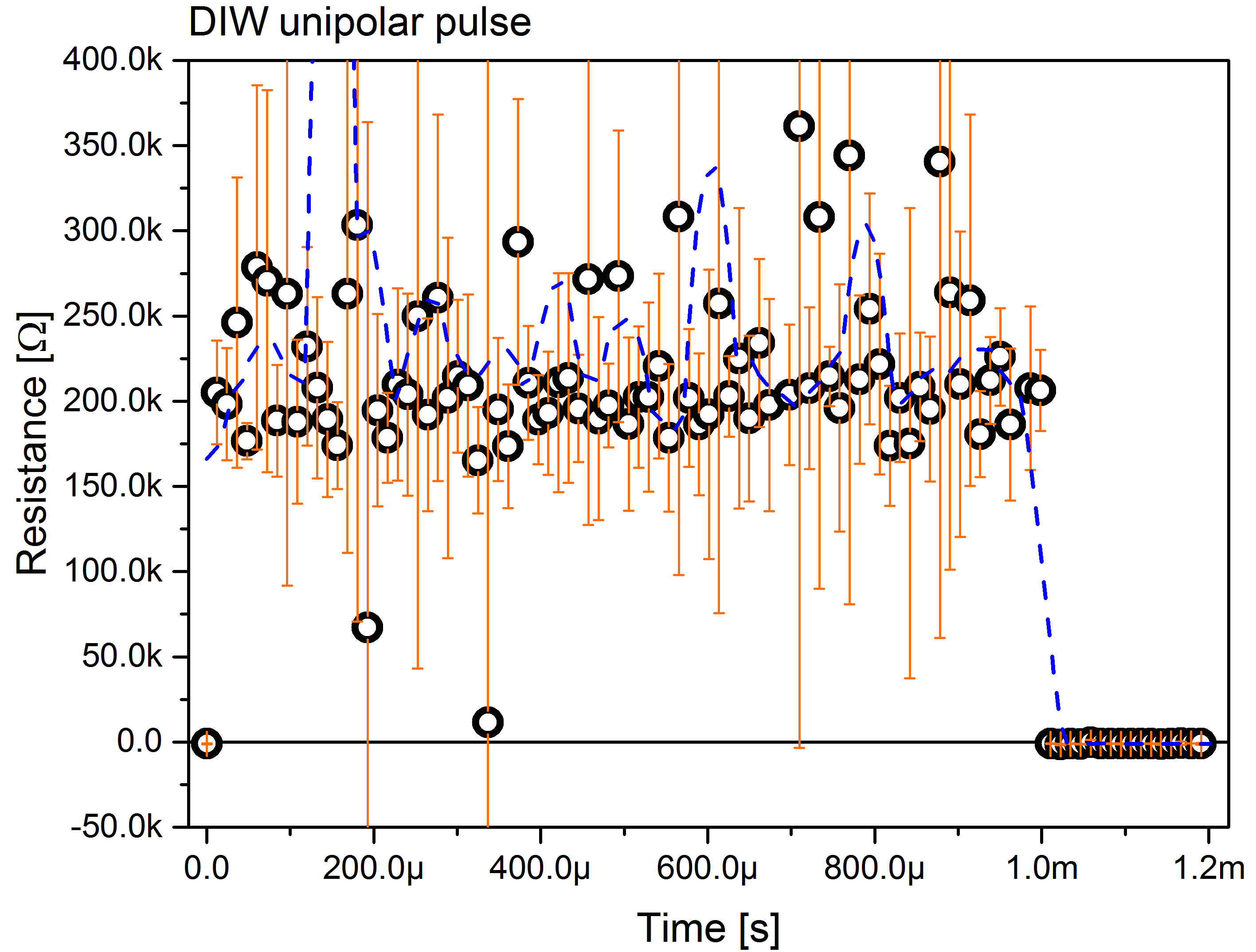}\label{fig19}}
    \caption{Unipolar pulse submitted to samples in (a)~the PT characterisation and (b)~DIW response to the same (dashed line: adjacent averaging over 50 points). 1 point of every 20 is shown for clarity.
    }
    \label{fig:18_19}
\end{figure}

PT pulse shape is presented in Fig.~\ref{fig:18_19}. DIW response to a unipolar pulse is shown, evidencing a noisy and rather flat resistance, with no distortion induced by high order harmonics; the rest resistance value is about -950~$\Omega$, that we may take as reference for further considerations on other materials.

\begin{figure}[htpb]
    \centering
    \subfigure[]{\includegraphics[scale=0.075]{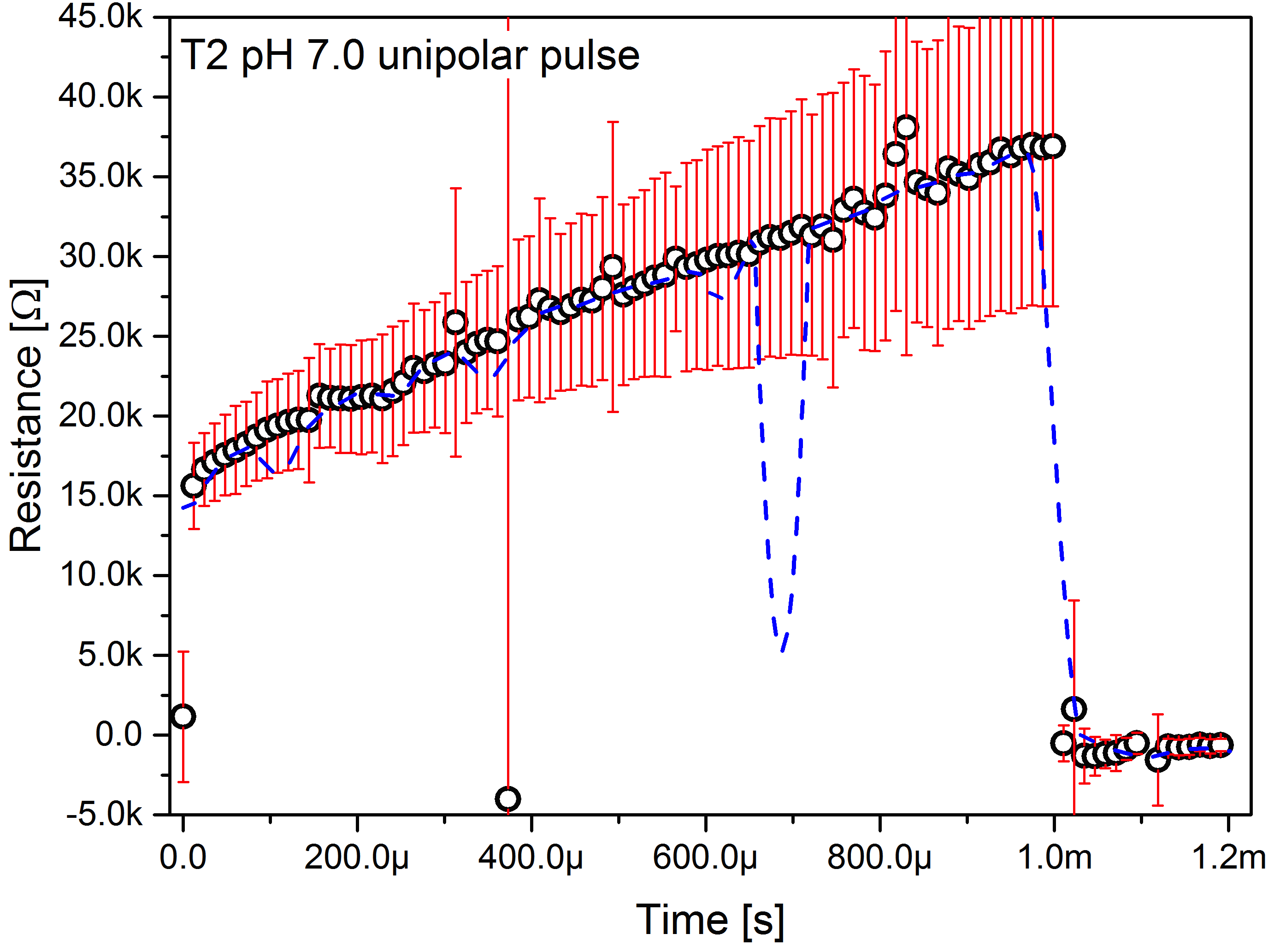}\label{fig20}}
    \subfigure[]{\includegraphics[scale=0.075]{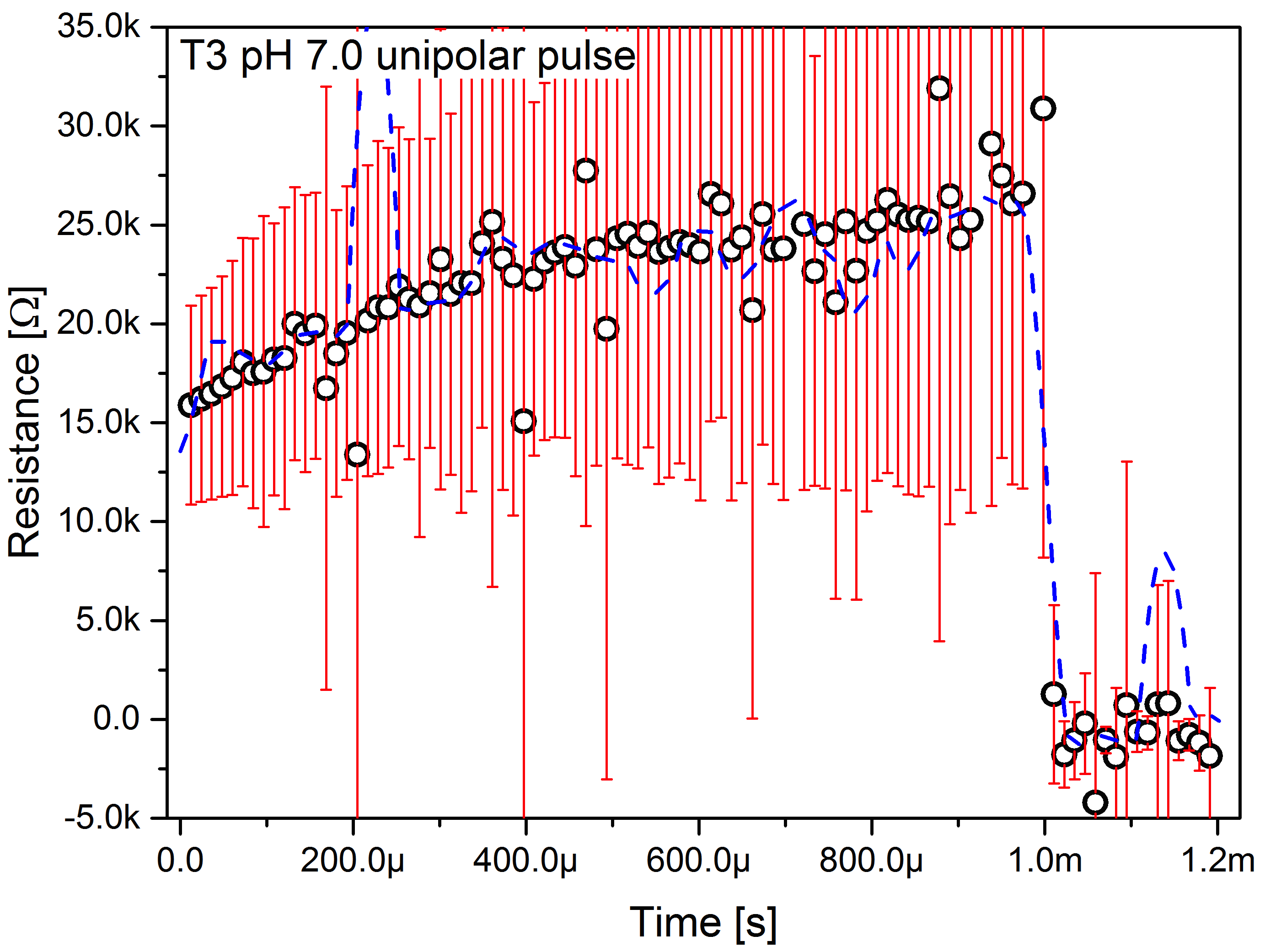}\label{fig21}}
    \subfigure[]{\includegraphics[scale=0.075]{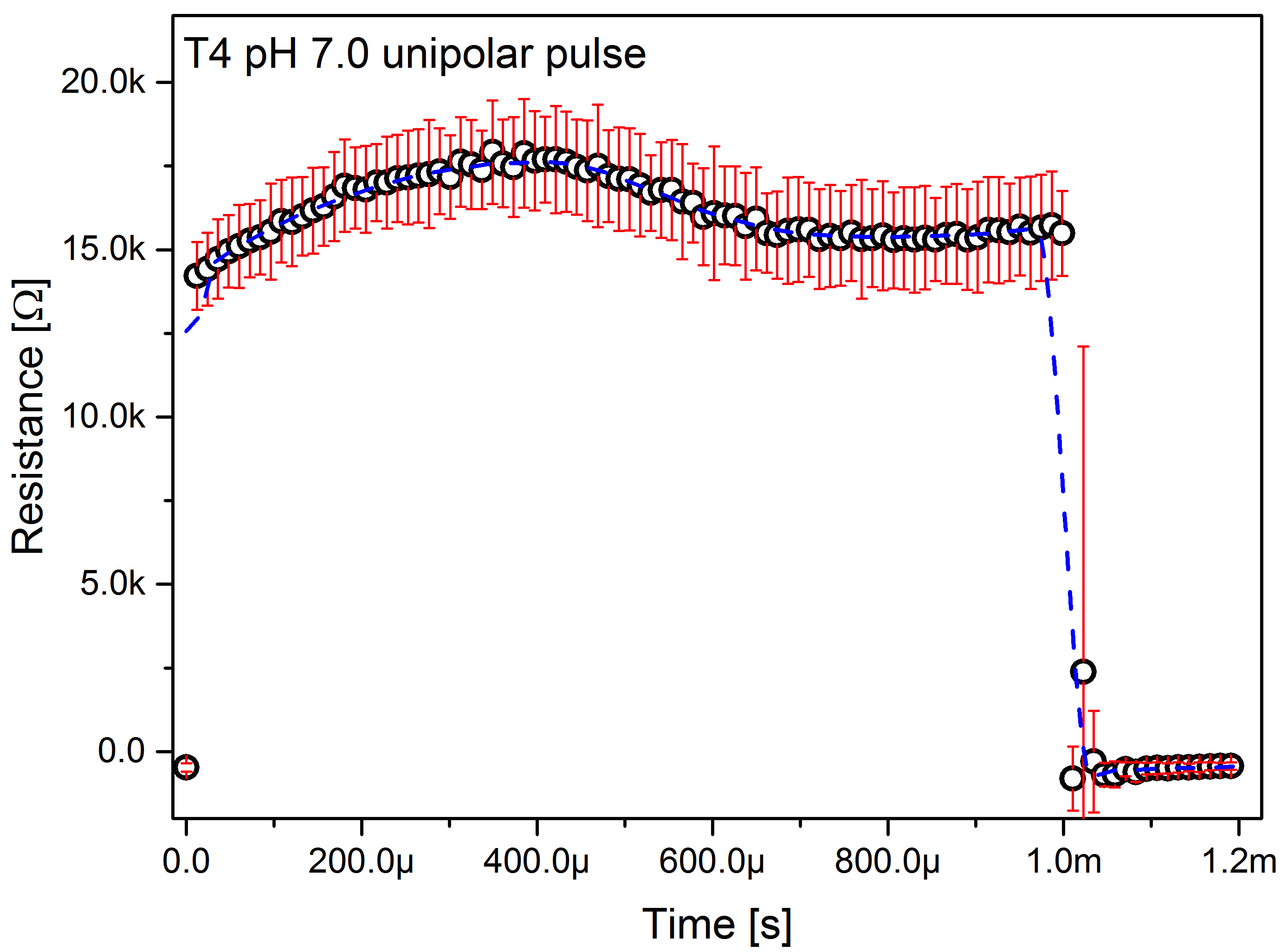}\label{fig22}}
      \subfigure[]{\includegraphics[scale=0.075]{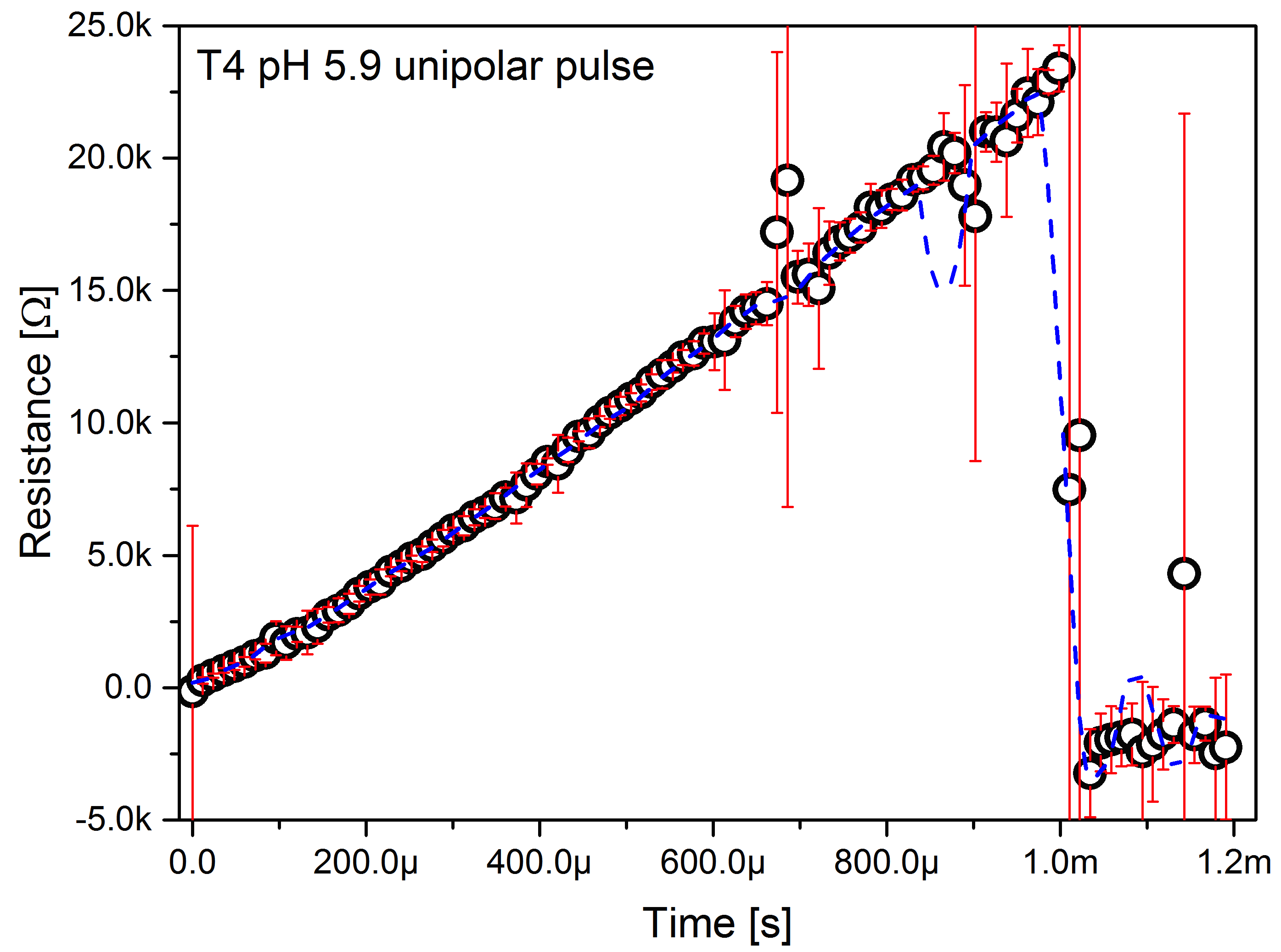}\label{fig23}}
        \subfigure[]{\includegraphics[scale=0.075]{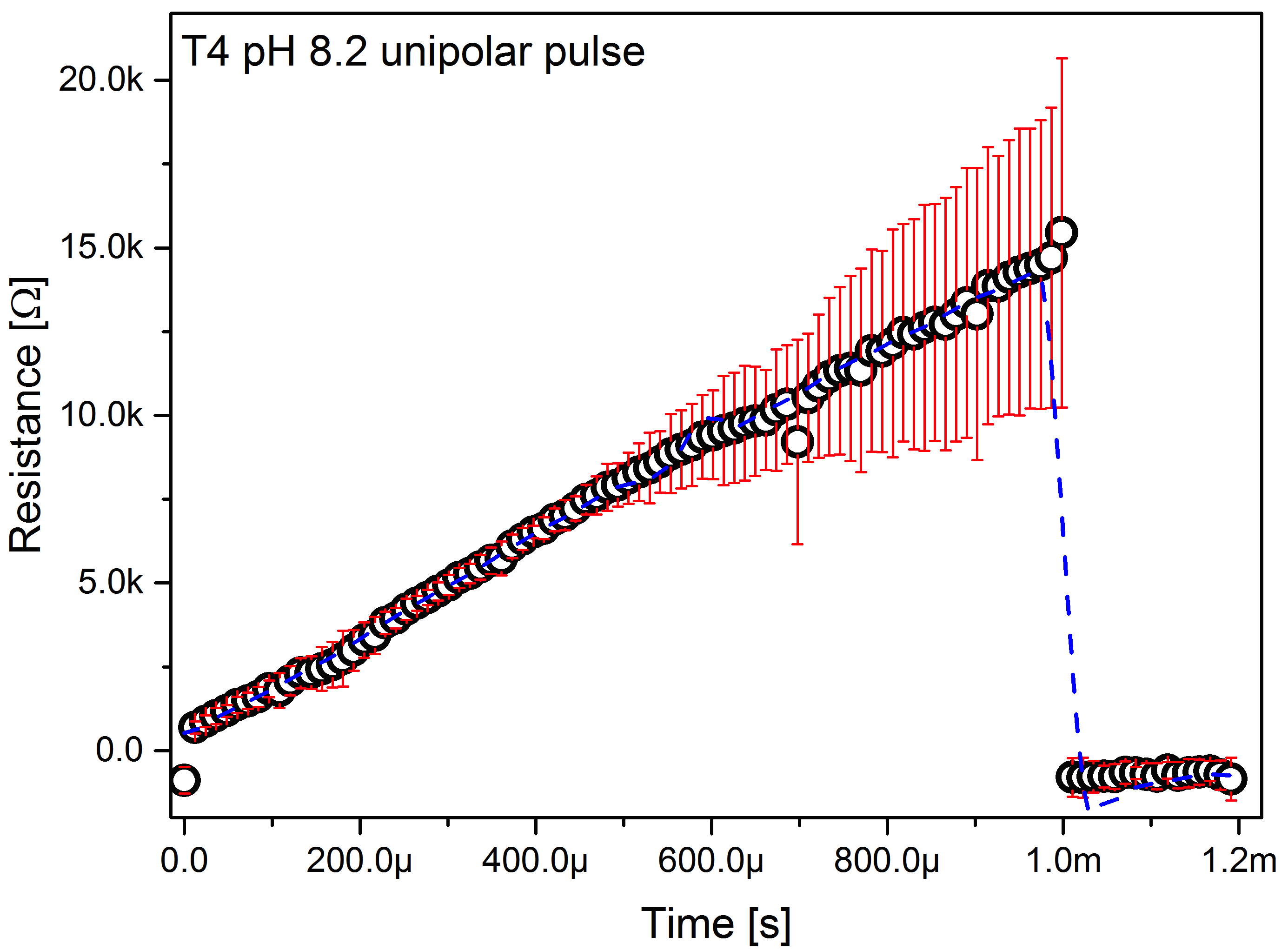}\label{fig24}}
    \caption{
    PT unipolar pulse submitted to T2 (a), T3 (b) and T4 (c) in neutral condition  and T4 in pH=5.9 (d) and T4 in pH=8.2 (e). Only 1 point every 10 shown for clarity. Average over 10 experimental curves (black open dots, red bars: standard deviation) superimposed to a smoothing over 50 adjacent points (dashed blue).
    }
    \label{fig:20_24}
\end{figure}

Shown in figures~\ref{fig20}--\ref{fig22} is a comparison between the unipolar pulse response of tectomers at neutral pH. The ultimate resistance achieved at the end of the positive waveform is inversely proportional to the number of legs of the tectomer. T4 shows a peculiar response, with a more stable profile and a much smaller deviation. The effect of pH is shown in  Figs.~\ref{fig23}--\ref{fig24}: both the acid and basic condition result in a linear response with very low deviation, at least in the first portion of the waveform. The noise pattern is different: the sample kept at pH=5.9 features random noise (Fig.~\ref{fig23}), while the sample at pH=8.2 features a slightly chaotic profile that in turn produces a high standard deviation over averaging (Fig.~\ref{fig24}).

\begin{figure}[htpb]
    \centering
    \subfigure[]{\includegraphics[scale=0.075]{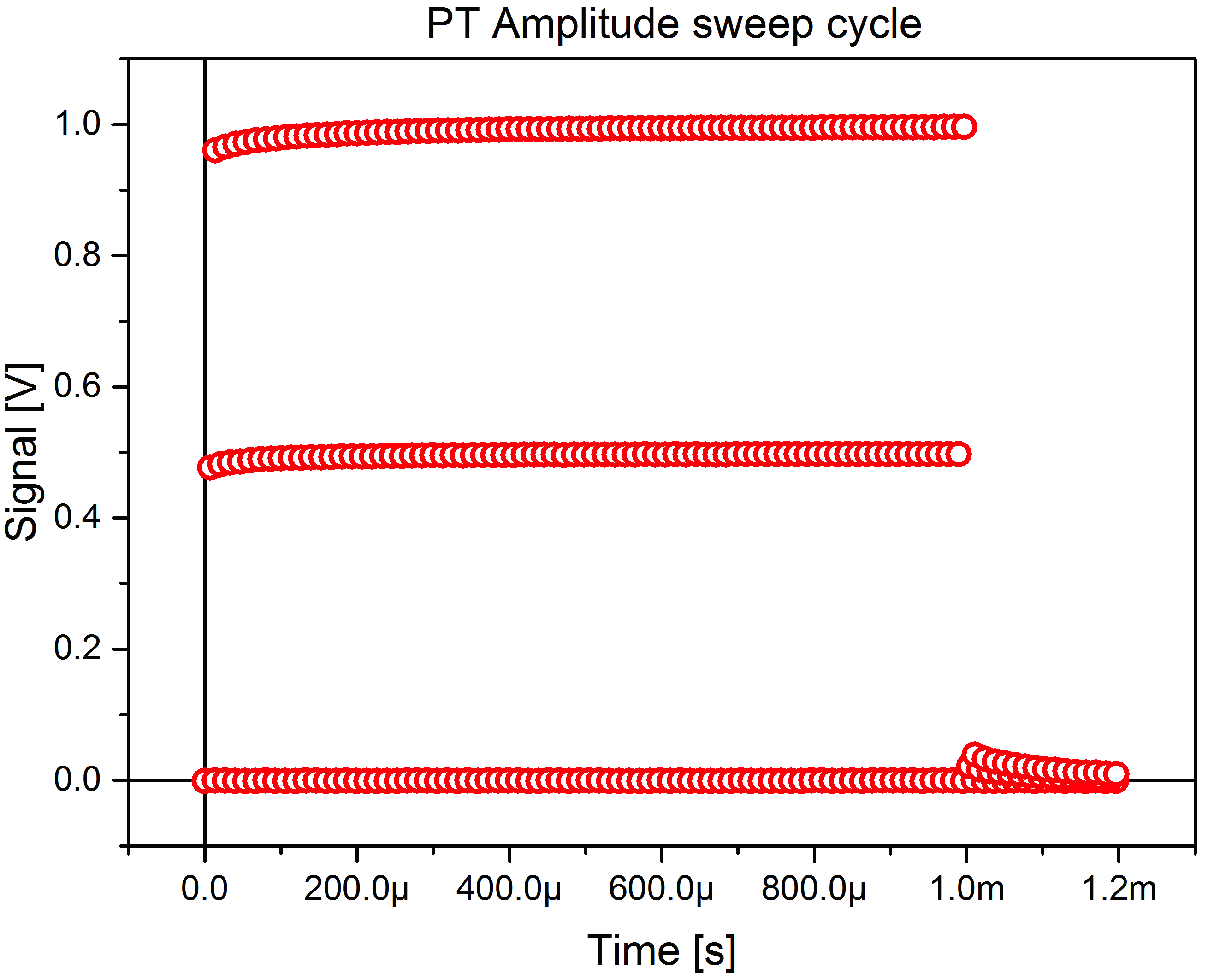}\label{fig25}}
    \subfigure[]{\includegraphics[scale=0.075]{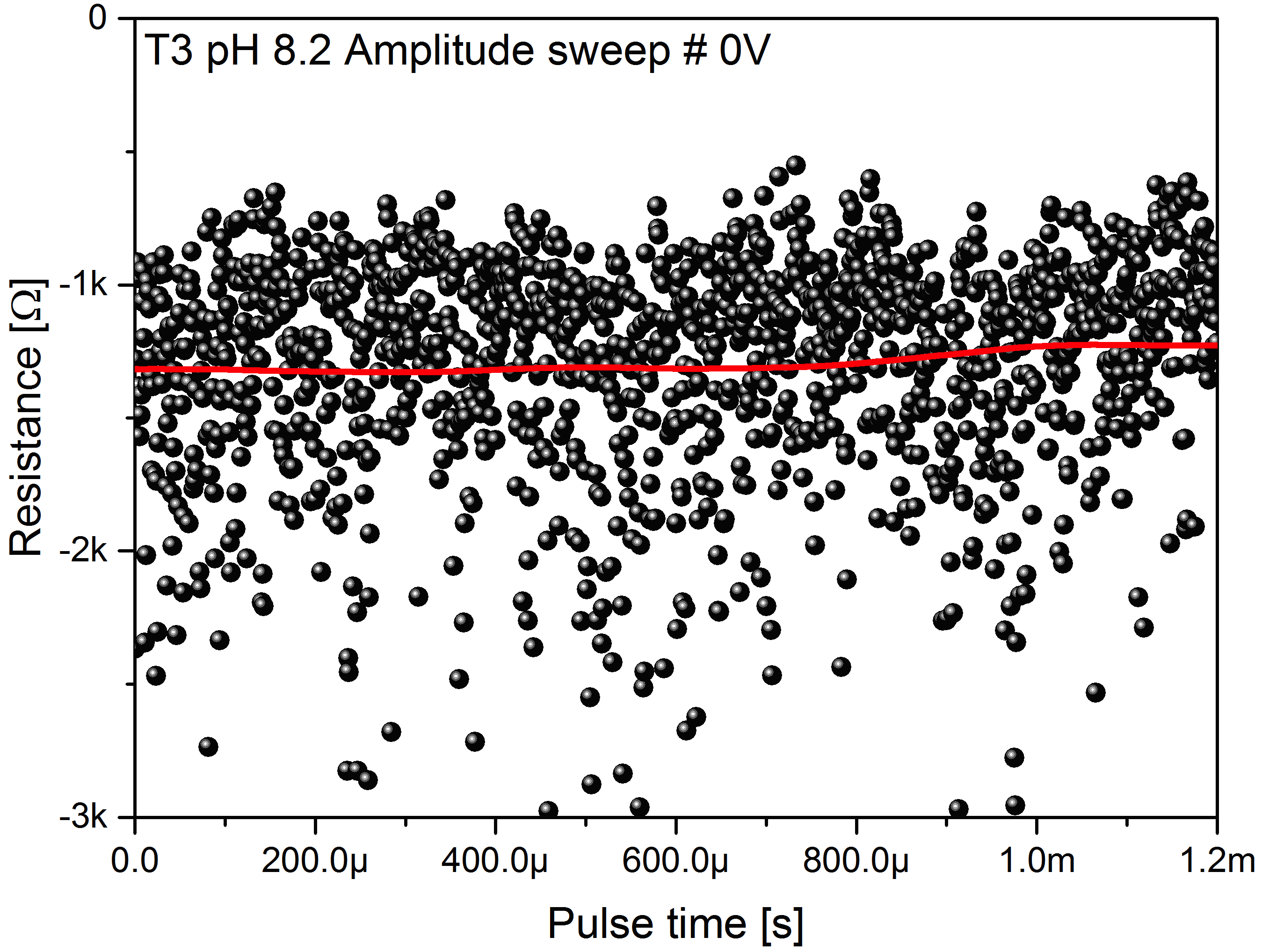}\label{fig26}}
    \subfigure[]{\includegraphics[scale=0.075]{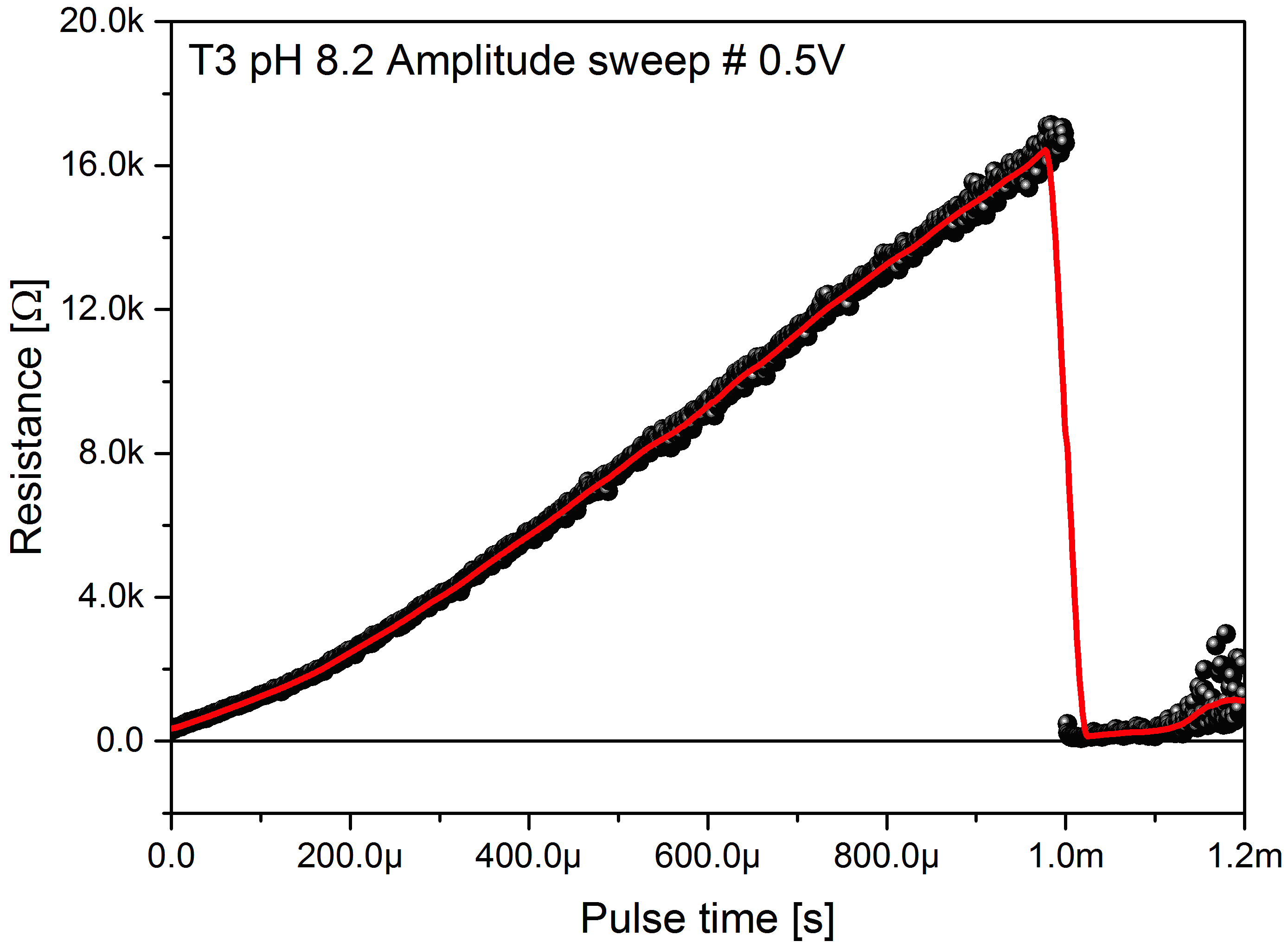}\label{fig27}}
      \subfigure[]{\includegraphics[scale=0.075]{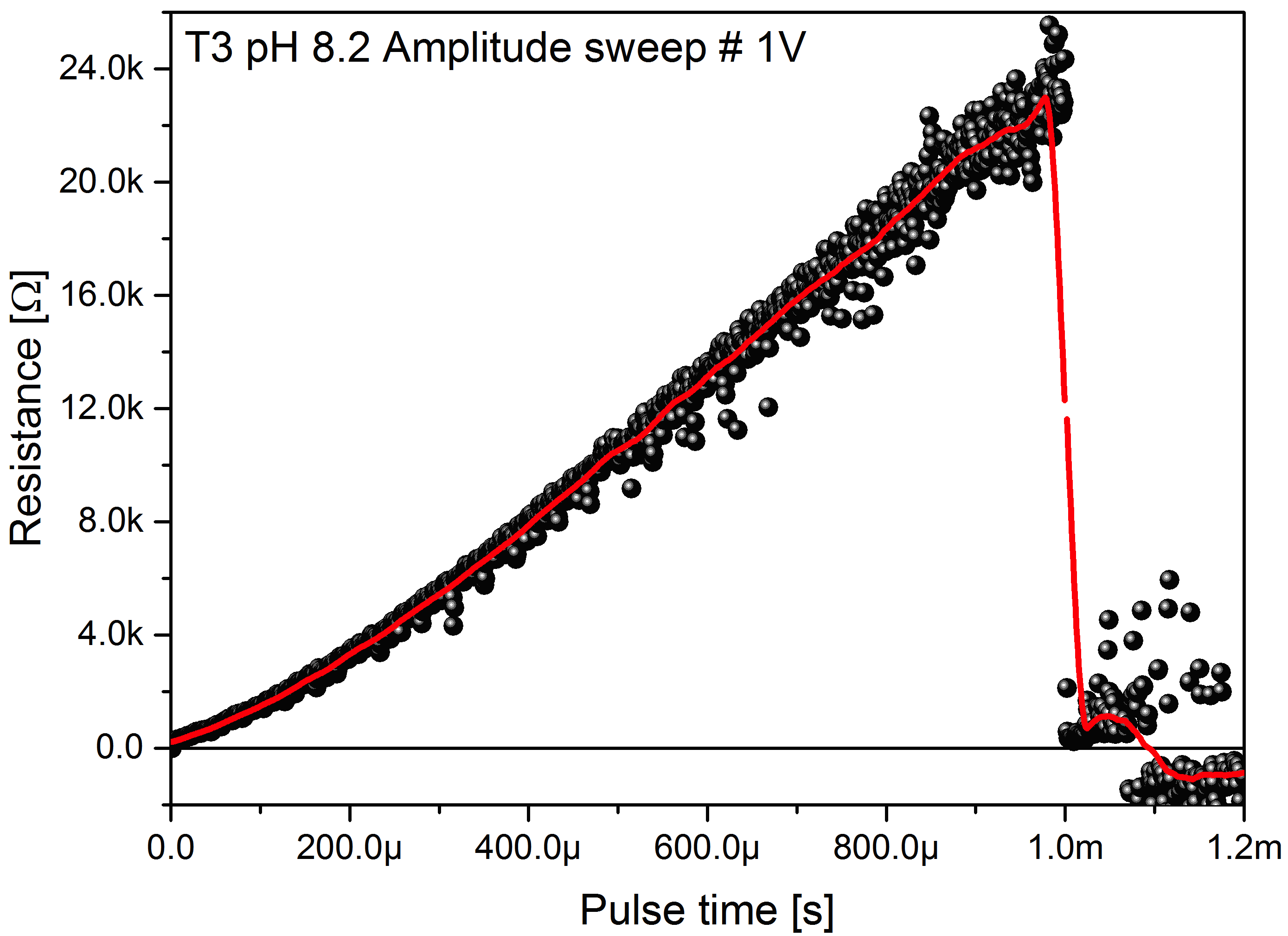}\label{fig28}}
    \caption{
 Amplitude sweep pulses submitted to samples in the PT characterisation (a), time-shifted to appear in the same graph. 1 point every 20 shown for clarity. Amplitude sweep response of sample T3 at pH=8.2 to the amplitude sweeps at 0~V (b), 0.5~V (c) and 1.0~V (d). All experimental points are shown.
    }
    \label{fig:25_28}
\end{figure}

In another operation mode, the amplitude sweep cycle, the sample is submitted to a train of pulses of growing amplitude, repeated 50 times. As shown in Fig.~\ref{fig:25_28}, we set three steps at 0, 0.5 and 1.0 V, each lasting 1~ms, and separated from the others by 4~ms of idle time at zero bias. Ten measurements are acquired as usual to follow any chaotic behaviour. Again in Fig.~\ref{fig:25_28} sample T3 at pH=8.2 has been shown, separating the three amplitude sweeps for clarity in different panels.

\subsection{Comparison}

To clarify the differences between tectomers and pH conditions, relevant data from the plots of both unipolar pulses and amplitude sweeps have been extracted and presented in Tab.~\ref{tab:1}. By comparing the slew rate (resistance increase speed, within the period of the pulse) extrapolated from the unipolar pulse measures and the amplitude sweep ones, we see a rather good agreement. The asymptotic resistance values at zero bias are nevertheless very different in the two techniques, and they show a positive correlation with the number of tails of the tectomer: the higher the number, the higher the absolute value of asymptotic terms, at any pH values. The slew rate shows on the contrary an inverse correlation with the number of tails, which holds in neutral conditions: the higher the number, the slower the resistance growth rate. When the solution is acid or basic, the fastest response is found either for T3 or T4. Interestingly, the highest slew rates are generally found for the intermediate potential (0.5~V) in the amplitude sweep condition. 

\begin{sidewaystable}[p]
    \centering
    \begin{tabular}{c||c|c|c|c|c|c|c|c}
Sample	&	R$\infty$@uni [$\Omega$]	&	Slew rate [$\Omega$/s]$\times 10^6$	&	R2 [-]	&	R$\infty$@ampl.sw [$\Omega$]	&	Slew rate @ 0.5 V [$\Omega$/s]$\times 10^7$	&	R2 [-]	&	Slew rate @ 1.0 V [$\Omega$/s]$\times 10^7$	&	R2 [-]	\\ \hline
T2 pH=7.0	&	-1095	&	(2.44$\Omega$0.03)$\times$10	&	0.95151	&	178168	&	(3.391$\pm$0.003)$\times$10	&	0.99951	&	(3.410$\pm$0.003)$\times$10	&	0.99946	\\ 
T3 pH=7.0	&	-81	&	(2.0$\pm$0.3)$\times$10	&	0.17399	&	235505	&	(3.54$\pm$0.04)$\times$10	&	0.88076	&	(2.30$\pm$0.02)$\times$10	&	0.96	\\ 
T4 pH=7.0	&	-436	&	(1.27$\pm$0.04)	&	0.88628	&	352601	&	(3.61$\pm$0.02)$\times$10	&	0.89996	&	(1.28$\pm$0.05)$\times$10	&	0.96	\\ 
T2 pH=5.9	&	-137	&	(2.00$\pm$0.01)$\times$10	&	0.99472	&	-739	&	(2.357$\pm$0.008)$\times$10	&	0.99591	&	(2.165$\pm$0.002)$\times$10	&	0.99975	\\ 
T3 pH=5.9	&	-1070	&	(4.57$\pm$0.01)$\times$10	&	0.99747	&	-991	&	(3.41$\pm$0.02)$\times$10	&	0.98763	&	(3.1$\pm$0.1)$\times$10	&	0.67551	\\ 
T4 pH=5.9	&	-1215	&	(2.352$\pm$0.004)$\times$10	&	0.99893	&	-2092	&	(2.962$\pm$0.006)$\times$10	&	0.99725	&	(2.653$\pm$0.007)$\times$10	&	0.99452	\\ 
T2 pH=8.2	&	-179	&	(2.850$\pm$0.003)$\times$10 &	0.99933	&	-908	&	(2.842$\pm$0.003)$\times$10	&	0.99929	&	(2.341$\pm$0.002)$\times$10	&	0.99928	\\ 
T3 pH=8.2	&	-1145	&	(2.224$\pm$0.003)$\times$10	&	0.9994	&	-1322	&	(1.866$\pm$0.002)$\times$10	&	0.9993	&	(2.641$\pm$0.002)$\times$10	&	0.99963	\\ 
T4 pH=8.2	&	-725	&	(1.543$\pm$0.001)$\times$10	&	0.99966	&	-2096	&	(2.96$\pm$0.03)$\times$10	&	0.94693	&	(3.05$\pm$0.04)$\times$10	&	0.92256	\\ 
    \end{tabular}
    \caption{Data extrapolated from the PT measurements, the first three columns in the unipolar pulse mode, the last five columns in the amplitude sweep mode. The asymptotic resistances R${\infty}$ are computed from the 50 points adjacent averaging smoothed curves, the slew rates are the slope of a linear fit to the smoothed curves, whose R2 is also reported.}
    \label{tab:1}
\end{sidewaystable}

\section{Summary}

We demonstrated that tectomers are versatile nanomaterials with non-trivial electronic properties which are, at least partly, determined by the pH of the medium and the number of olygoglycine tails. Key findings are as follows:
\begin{itemize}
    \item Direct current-voltage (IV) cycles measured on 4-tailed tectomers at neutral pH show increase of the maximum current flowing with each cycle, more likely due to re-ordering of tectomers and formation of layers with increased conductivity.
    \item  In some samples, oscillations of IV curve have been observed, which might indicate that in some parts of the droplets assembly and disassembly processes co-exist in the same droplet.
    \item Behaviour of the tectomers during IV swipe is determined by acidity of the medium and the number of tails (2,3, or 4). The IV curves show that 2-tailed species in acidic environment (amorphous case) can be used as resistive switching devices.
    \item   We found that it is possible to distinguish between the different tectomers and/or the pH of their solution by measuring their impedance in the range between 1~kHz and 1~MHz.
    \item Three-tailed tectomers could be used as voltage controlled gain passive devices, because their hysteretic curve show similar reactances but opposite sign resistances. 
    \item Three-tailed tectomers show ferroelectric-like response, under acidic solution (amorphous case).
\item  Asymptotic resistance of tectomers is proportional to a number of tails while resistance increase speed during pulse stimulation is inversely proportional to a number of tails.
\end{itemize}

\FloatBarrier
\section{Supplementary information}

\begin{figure}[htbp]
    \centering
    \subfigure[]{\includegraphics[scale=0.075]{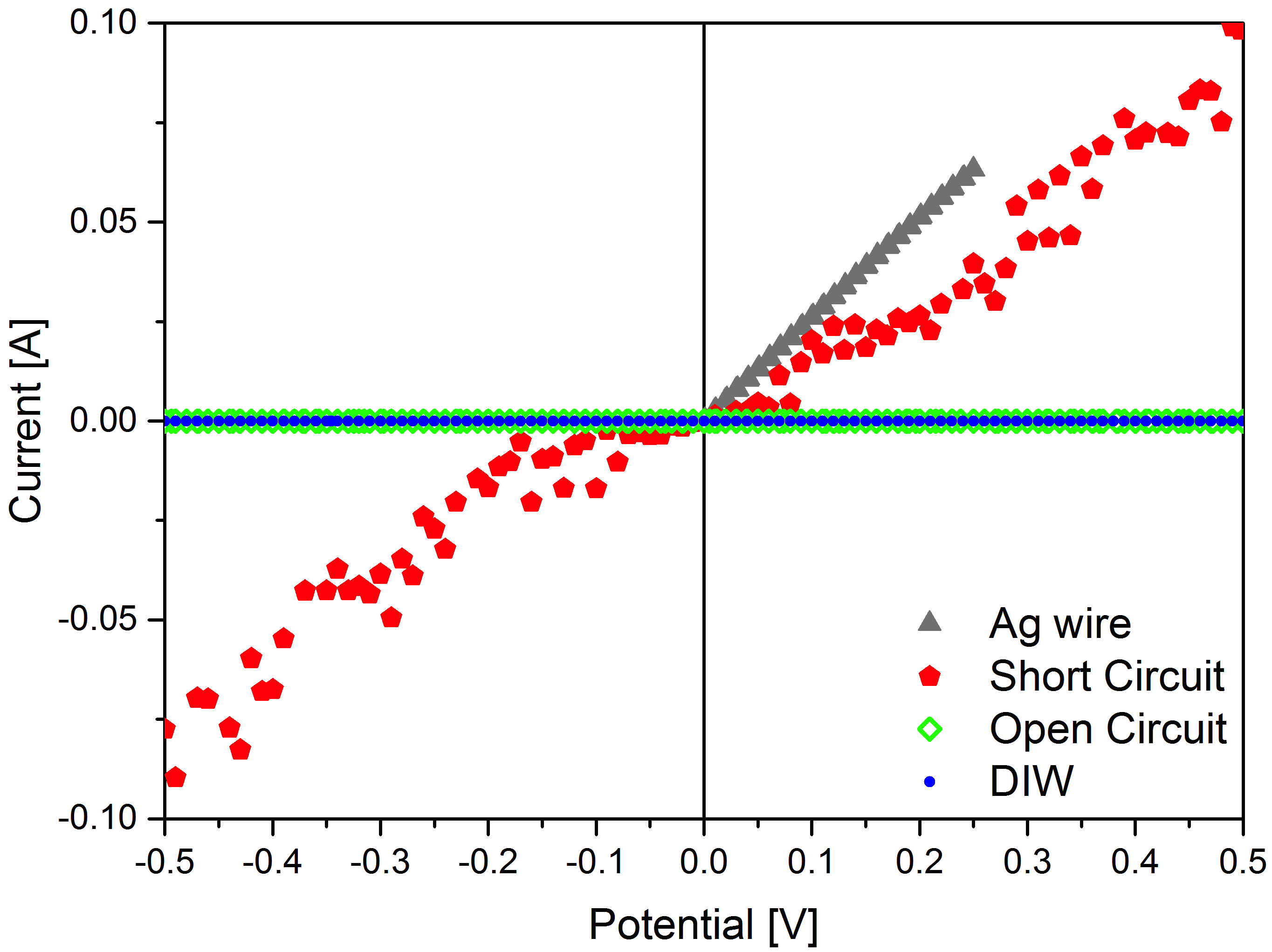}\label{figS1}}
    \subfigure[]{\includegraphics[scale=0.075]{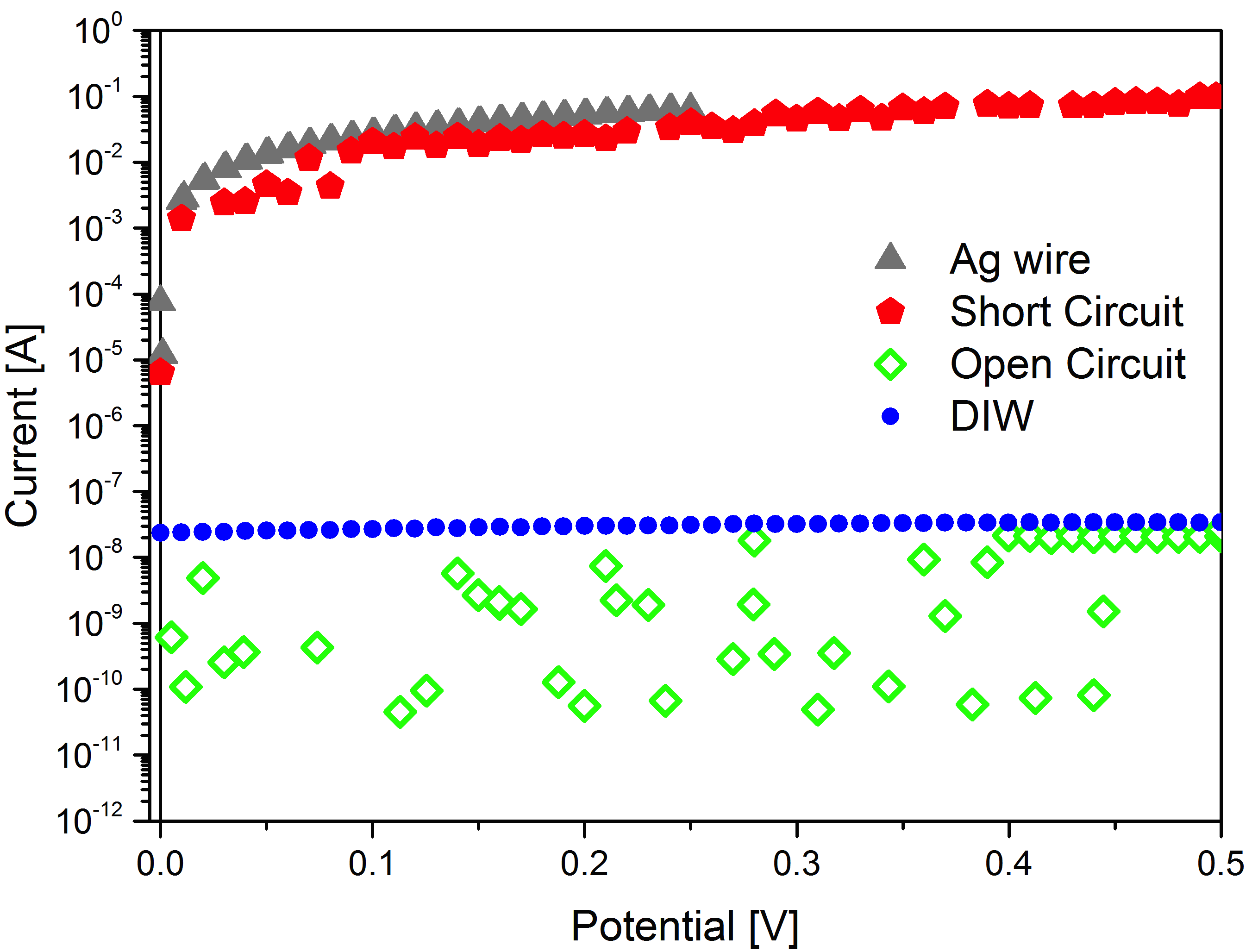}\label{figS2}}
    \subfigure[]{\includegraphics[scale=0.075]{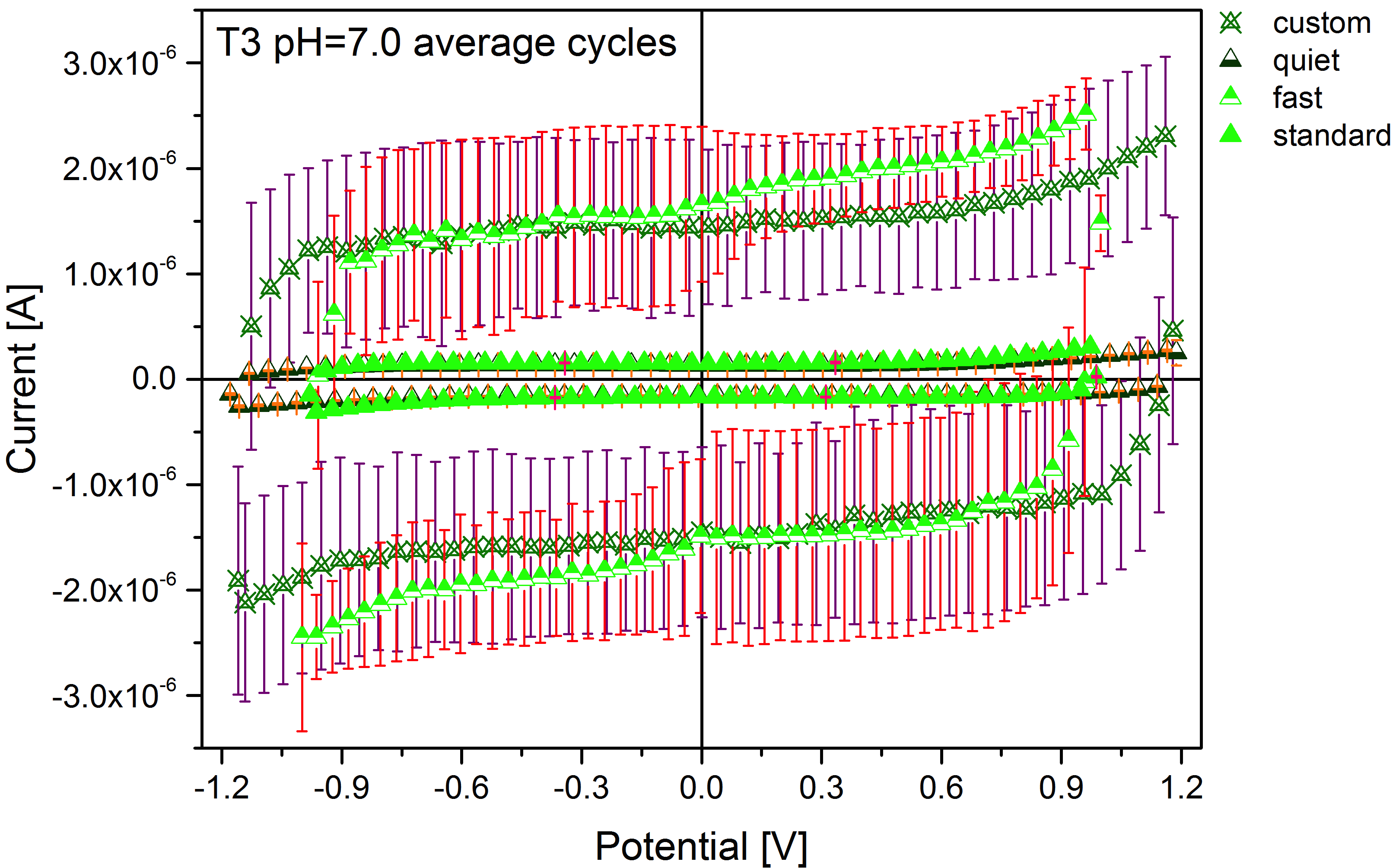}\label{figS3}}
    \caption{
    (a)~IV curves of a pure Ag wire (Ohmic), of a short circuit (Ohmic with noise) and of a droplet of DIW. 1 point every 10 is shown for clarity.
    (b)~IV curves of a pure Ag wire (Ohmic) of an open circuit and of a droplet of DIW. 1 point every 10 is shown for clarity.
    (c)~Average IV curves of a solution of T3 in DIW, neutral pH, measured with four different timings. 1 point every 10 is shown for clarity.
    }
    \label{fig:S1S2S3}
\end{figure}

Here we report the measurements taken on blank samples/other electrically relevant conditions such as short circuit and open circuit (Fig.~\ref{figS1}). The maximum current is clamped at 100~mA. Both the Ag wire and the short circuit show an Ohmic response, though in the second case a higher noise level is found, due to the electrodes. To see a difference between the open circuit and the DIW droplet, refer to the logarithmic scale of Fig.~\ref{figS2}. The current level is on average 5~nA for the open circuit and 30~nA for DIW. The effect of measurement timing was also assessed, in particular considering that each IV cycle is completed in about 10~s, we may consider that DC measurements are rather low frequency AC ones. Therefore the number of real measures taken to produce one single experimental point do influence the response, eventually triggering the capacitive component for the shortest and fastest measures. In Fig.~\ref{figS3} a comparison between 4 different timings during 10 subsequent DC cycles measured on a droplet of T3 at pH~7.0 is shown: quiet (slowest), standard (intermediate), custom (intermediate fast) and fast. Both the standard and quiet measures have very narrow standard deviations and a thin hysteresis consistent with a low capacitive component triggered. Both the custom and fast measures have broad error bands and a much higher hysteresis.

\begin{figure}[htbp]
    \centering
    \subfigure[]{\includegraphics[scale=0.075]{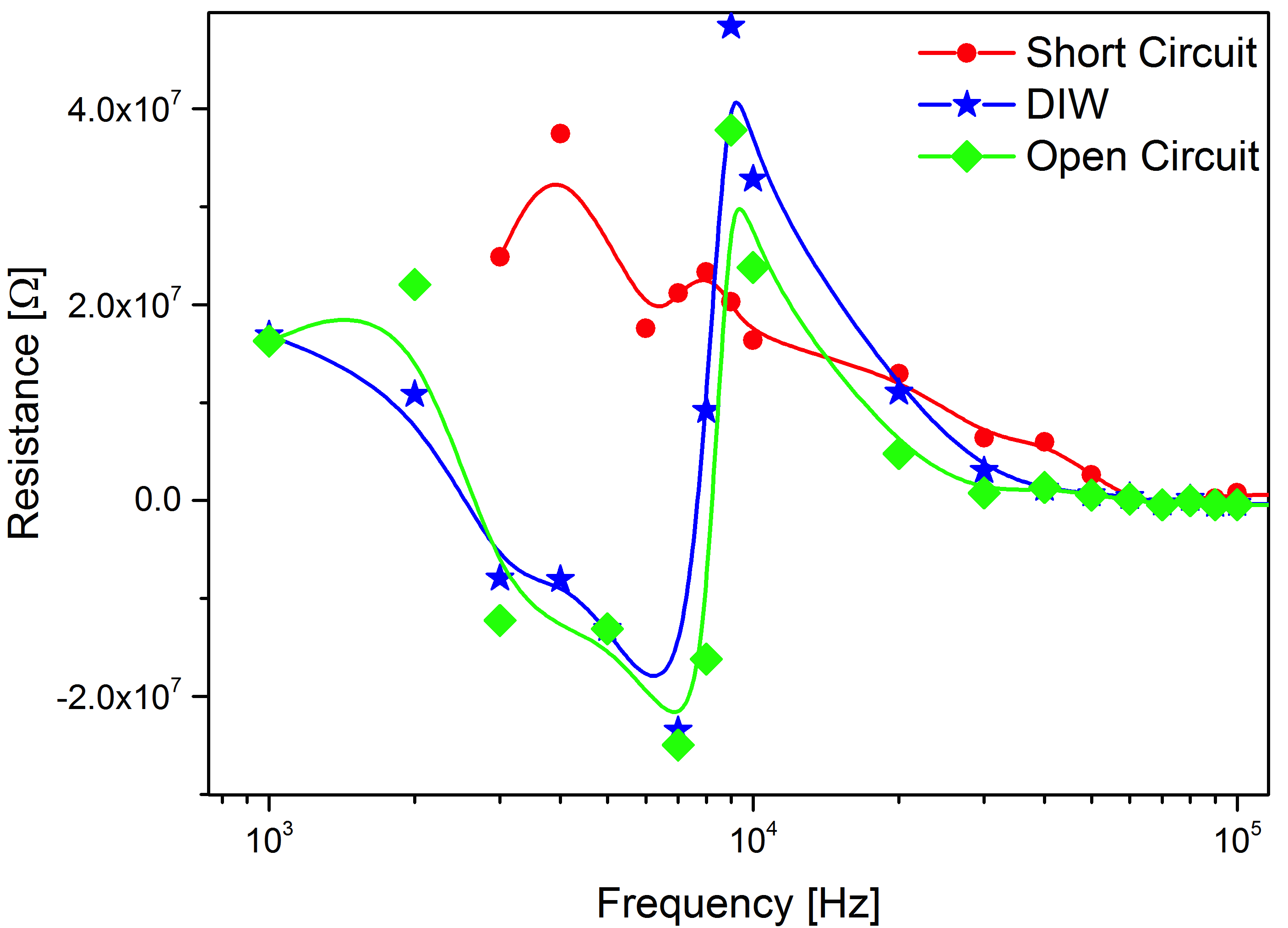}\label{figS4a}}
    \subfigure[]{\includegraphics[scale=0.075]{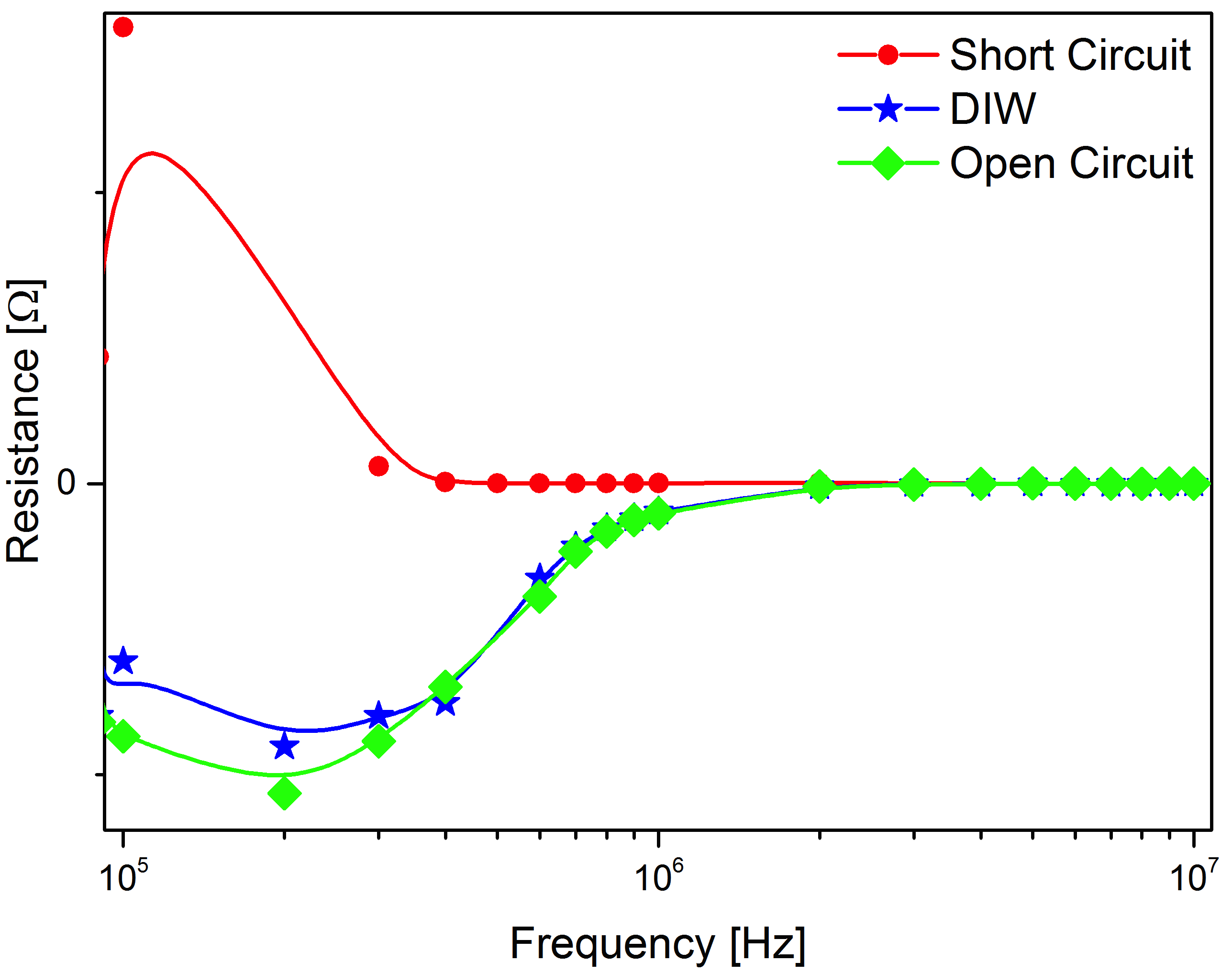}\label{figS4b}}
    \subfigure[]{\includegraphics[scale=0.075]{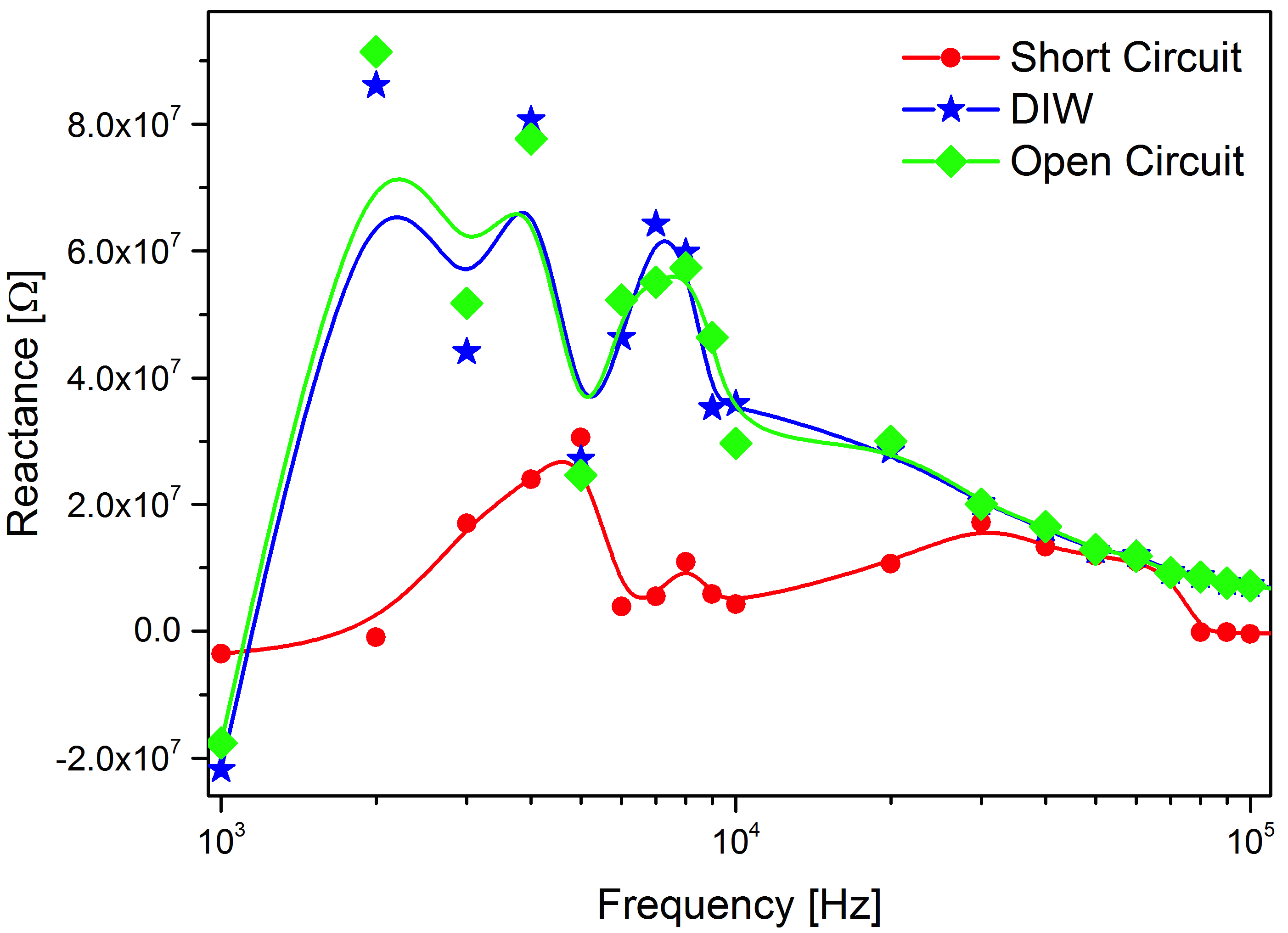}\label{figS4c}}
    \subfigure[]{\includegraphics[scale=0.075]{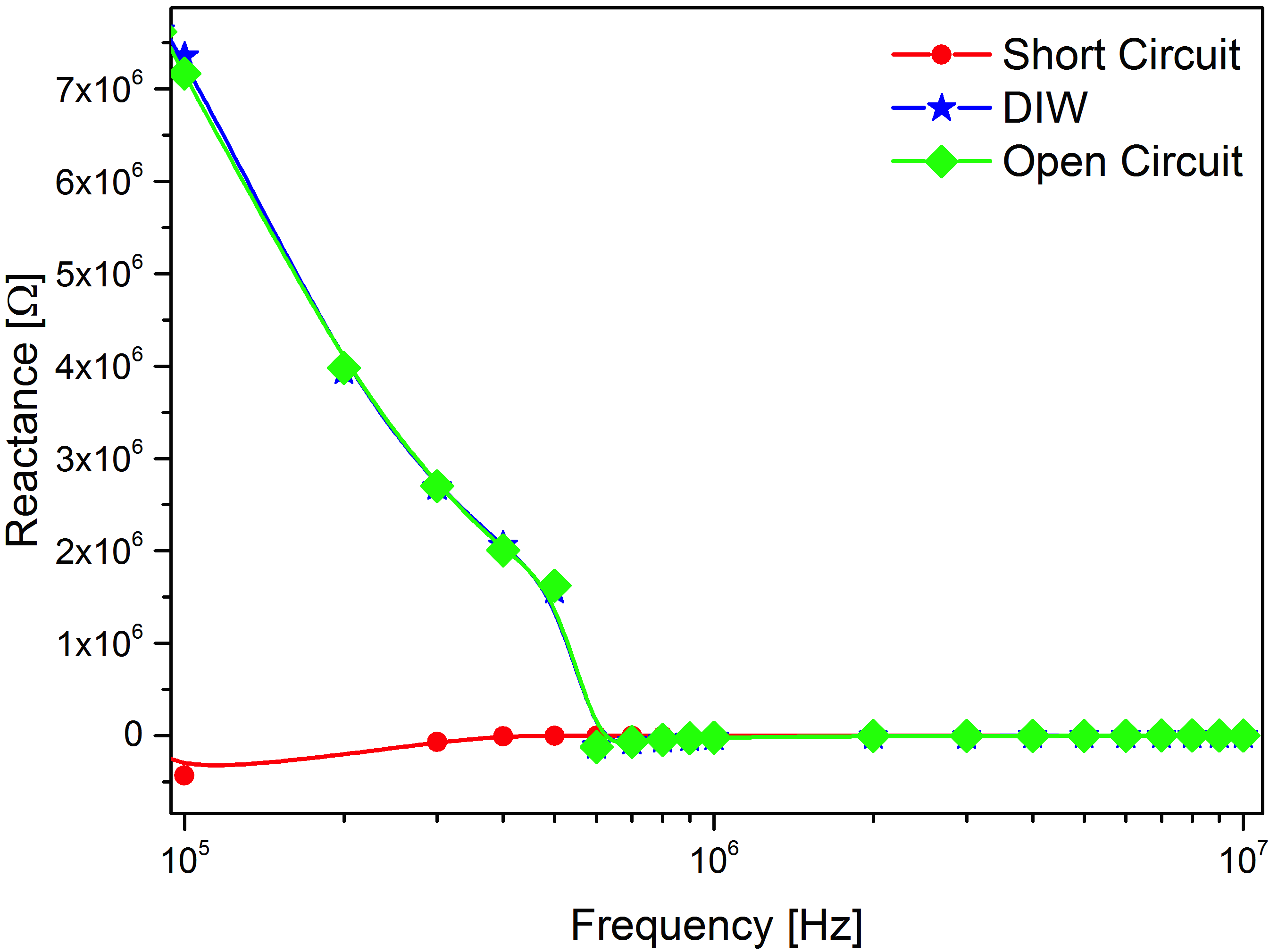}\label{figS4d}}
    \caption{
    Frequency sweeps showing the impedance of a short circuit, a DIW water and an open circuit condition: resistance (a) and reactance (b), low frequency regime (c) and high frequency regime (d).
    }
    \label{fig:S4}
\end{figure}

Blank measurements in the AC mode have been performed on DIW, short circuit and open circuit. The response is shown in Fig.~\ref{fig:S4} dividing the two relevant frequency ranges in low (between 1~kHz and 100~kHz) and high (between 100~kHz and 10~MHz), and dividing complex impedance in resistance and reactance. Resistance plots show that it is not so easy to discriminate between DIW and open circuit, they both present a peculiar Fuchsian discontinuity around 8~kHz where resistance changes from negative to positive. This is expected, due to the very high resistivity of DIW. Above 2~MHz it's also difficult to discriminate open circuit from short circuit because the leads become antennas. Reactance plots also show that DIW and open circuit have the same behaviour, confirming that the connecting leads have a bad performance when the DUT has a high impedance.

\begin{figure}[htpb]
    \centering
    \subfigure[]{\includegraphics[scale=0.075]{Reportontectomers-img020}\label{figS5a}}
    \subfigure[]{\includegraphics[scale=0.075]{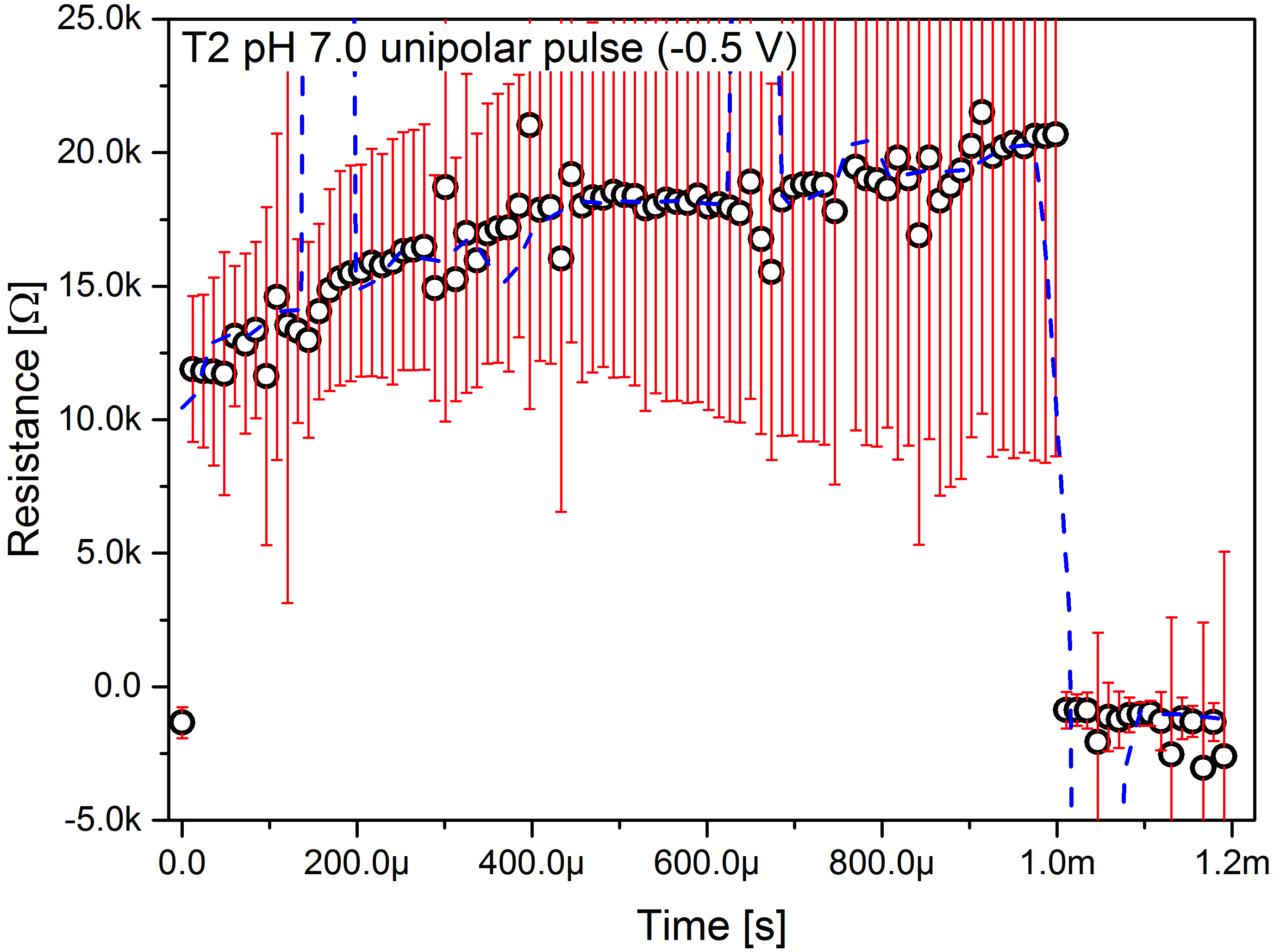}\label{figS5b}}
    \caption{
PT unipolar pulse submitted to T2 in neutral condition, with a positive (a) and a negative (b) pulse of 0.5~V amplitude. Only 1 point every 10 shown for clarity. Average over 10 experimental curves (black open dots, red bars: standard deviation) superimposed to a smoothing over 50 adjacent points (dashed blue).
    }
    \label{fig:S5}
\end{figure}

PT measurements with unipolar pulses have been characterised using different pulse amplitude. It was found that the pulse amplitude has a strong influence in the response of the tectomers. Fig.~\ref{fig:S5} shows the responses of T2 in neutral condition submitted to pulses of the same absolute values but different signs. The negative pulse results in a lower typical resistance. The noise in the average curves is due to a chaotic behaviour which is found in both cases.

\FloatBarrier
\bibliographystyle{plain}
\bibliography{bibliography}

\end{document}